\newif\ifprocs
\procsfalse

\newcommand{\procsonly}[1]{\ifprocs#1\fi}
\newcommand{\fullonly}[1]{\ifprocs\else#1\fi}

\ifprocs
\documentclass[oribibl]{llncs}
\else
\documentclass[11pt,letterpaper]{article}
\usepackage{typearea}
\typearea{15}
\usepackage[compact]{titlesec}

\usepackage{theorem,latexsym,graphicx}
\usepackage{bm}
\fi
\usepackage{ifpdf}
\usepackage{amsmath,amssymb,enumerate}
\usepackage{shadow,shadethm,color}
\usepackage{algorithm, algorithmic}
\usepackage{array,tabularx,mdwtab}
\usepackage{xspace}

\usepackage{amsmath,amssymb,array,csquotes,units}

\ifprocs
\makeatletter
\renewenvironment{abstract}{
      \list{}{\advance\topsep by0.35cm\relax\small
      \leftmargin=1cm
      \labelwidth=\z@
      \listparindent=1.5em
      \itemindent\listparindent
      \rightmargin\leftmargin}\item[\hskip\labelsep\hskip-\itemindent
                                    \bfseries\abstractname]}
    {\endlist}
\makeatother
\fi

\definecolor{Darkblue}{rgb}{0,0,0.4}
\definecolor{Brown}{cmyk}{0,0.81,1.,0.60}
\definecolor{Purple}{cmyk}{0.45,0.86,0,0}
\newcommand{\mydriver}{hypertex}
\ifpdf
 \renewcommand{\mydriver}{pdftex}
\fi
\usepackage[breaklinks,\mydriver]{hyperref}
\hypersetup{colorlinks=true,
            citebordercolor={.6 .6 .6},linkbordercolor={.6 .6 .6},
citecolor=Darkblue,urlcolor=black,linkcolor=Darkblue}
\newcommand{\lref}[2][]{\hyperref[#2]{#1~\ref*{#2}}}

\ifprocs
\else
\fi

\ifprocs\else
\newtheorem{theorem}{Theorem}[section]

\newtheorem{lemma}[theorem]{Lemma}
\newshadetheorem{lemmashaded}[theorem]{Lemma}

\newtheorem{claim}[theorem]{Claim}

\newtheorem{corollary}[theorem]{Corollary}

\newenvironment{proof}{

\noindent{\bf Proof:}}
{\hfill$\blacksquare$

}

\fi

\newcommand{\junk}[1]{}
\newcommand{\ignore}[1]{}

\newcommand{\R}[0]{{\ensuremath{\mathbb{R}}}}
\newcommand{\Rplus}[0]{{\ensuremath{\mathbb{R}_{\geq 0}}}}

\newcommand{\Z}[0]{{\ensuremath{\mathbb{Z}}}}

\newcommand{\sse}{\subseteq}
\newcommand{\A}{{\mathcal{A}}}

\newcommand{\M}{{\mathcal{M}}}
\newcommand{\Mat}{{\mathbf{M}}}
\renewcommand{\P}{{\mathcal{P}}}

\newcommand{\eps}{\varepsilon}

\newcommand{\zex}{{$0$-extension}\xspace}
\ifprocs
\renewcommand{\ts}{\textstyle}
\else
\newcommand{\ts}{\textstyle}
\fi

\DeclareMathOperator{\EX}{{\mathbb E}}
\DeclareMathOperator{\diam}{diam}
\DeclareMathOperator{\supp}{supp}

\newcommand{\alphafhrt}{\smash{O(\frac{\log k}{\log \log k})}}
\newcommand{\Racke}{R{\"a}cke}

\ifprocs\else
\newcounter{note}[section]

\fi

\newcommand{\remove}[1]{}
\def\compactify{\itemsep=0pt \topsep=0pt \partopsep=0pt \parsep=0pt}

\newcommand{\initOneLiners}{
    \setlength{\itemsep}{0pt}
    \setlength{\parsep }{0pt}
    \setlength{\topsep }{0pt}
}
\newenvironment{OneLiners}[1][\ensuremath{\bullet}]
    {\begin{list}
        {#1}
        {\initOneLiners}}
    {\end{list}}

\newcommand{\Otilde}{{\widetilde{O}}}

\begin{document}

\title{\texorpdfstring{Vertex Sparsifiers: New Results from Old Techniques
\thanks{A preliminary version appeared in the Proceedings of the 13th Workshop on Approximation Algorithms for Combinatorial Optimization Problems (APPROX 2010).}
}{Vertex Sparsifiers: New Results from Old Techniques}
}

\ifprocs
\else
\author{
  Matthias Englert\thanks{Department of Computer Science and DIMAP, University of Warwick,
   Coventry, UK.  Supported by EPSRC grant EP/F043333/1 and DIMAP (the Centre for Discrete
   Mathematics and its Applications). \texttt{englert@dcs.warwick.ac.uk}.}
  \\University of Warwick
  \and
  Anupam Gupta\thanks{Computer Science Department, Carnegie Mellon
    University, Pittsburgh, PA 15213, USA. Research was partly supported by
    the NSF award CCF-0729022, and an Alfred P.~Sloan Fellowship. This
    research was done when visiting Microsoft Research SVC, La Avenida,
    Mountain View CA. \texttt{anupamg@cs.cmu.edu}.}
  \\Carnegie Mellon
  \and
  Robert Krauthgamer
  \thanks{This work was supported in part by The Israel Science Foundation
    (grant \#452/08), and by a Minerva grant.
    Weizmann Institute of Science, Rehovot, Israel.
    \texttt{robert.krauthgamer@weizmann.ac.il}.
  }
  \\Weizmann Institute
  \and
  Harald R\"{a}cke\thanks{Institut f\"{u}r Informatik, Technische Universit\"{a}t M\"{u}nchen,
   Munich, Germany. \texttt{raecke@in.tum.de}.}
  \\TU M\"{u}nchen
  \and
  Inbal Talgam-Cohen\thanks{Computer Science Department, Stanford University, Stanford, CA 94350, USA.}
  \\Stanford University
  \and
  Kunal Talwar\thanks{Microsoft Research Silicon Valley, 1065 La Avenida, Mountain View, CA, USA. \texttt{kunal@microsoft.com}.}
  \\MSR SVC
}
\fi

\ifprocs\else
\begin{titlepage}
  \def\thepage{}
  \thispagestyle{empty}
\fi

  \maketitle

  \begin{abstract}

    Given a capacitated graph $G = (V,E)$ and a set of terminals $K \sse
    V$, how should we produce a graph $H$ only on the terminals $K$ so
    that every (multicommodity) flow between the terminals in $G$ could
    be supported in $H$ with low congestion, and vice versa? (Such a
    graph $H$ is called a \emph{flow-sparsifier} for $G$.)  What if we
    want $H$ to be a ``simple'' graph?  What if we allow $H$ to be a
    convex combination of simple graphs? 

    Improving on results of Moitra [FOCS 2009]
    and Leighton and Moitra [STOC 2010], we give efficient algorithms for
    constructing: (a) a flow-sparsifier $H$ that maintains 
    congestion
    up to a factor of $\alphafhrt$, where $k = |K|$;

    (b) a convex combination
    of trees over the terminals $K$ that maintains congestion up to a
    factor of $O(\log k)$; (c) for a planar graph $G$, a convex combination of
    planar graphs that maintains congestion
    up to a constant factor.  This requires us to give a new algorithm
    for the \zex problem, the first one in which the preimages of each
    terminal are connected in $G$.  Moreover, this result extends to
    minor-closed families of graphs.

    Our bounds immediately imply improved approximation
    guarantees for several terminal-based cut and ordering problems.
  \end{abstract}

\ifprocs\else
\end{titlepage}
\newpage
\fi

\section{Introduction}
\label{sec:introduction}

\noindent
Given an undirected capacitated graph $G = (V,E)$ and a set of terminal
nodes $K \sse V$, we consider the question of producing a graph $H$ only
on the terminals $K$ so that the congestion incurred on $G$ and $H$ for
any multicommodity flow routed between terminal nodes is similar.
Often, we will want the graph $H$ to be structurally ``simpler'' than
$G$ as well. Such a graph $H$ will be called a \emph{flow-sparsifier}
for $G$; the \emph{loss} (also known as \emph{quality}) of the flow-sparsifier is the factor by which
the congestions in the graphs $G$ and $H$ differ. For instance, when $K
= V$, the results of \Racke~\cite{R08} give a convex combination of trees $H$
with a loss of $O(\log n)$. We call this a \emph{tree-based
  flow-sparsifier}, meaning that it is a convex combination of
trees.\footnote{More generally, for a class $\mathcal{F}$ of graphs, we define an
  $\mathcal{F}$-flow-sparsifier to be a sparsifier that uses a single
  graph from $\mathcal{F}$, and an $\mathcal{F}$-based flow-sparsifier to
  be a sparsifier that uses a convex combination of graphs from
  $\mathcal{F}$.} 
Here and throughout, $k = |K|$ denotes the number of
terminals, and $n=|V|$ the size of the graph.

For the case where $K \neq V$, it was shown by Moitra~\cite{Moitra09}
and by Leighton and Moitra~\cite{LM10} that for every $G$ and $K$, there
exists a flow-sparsifier $H = (K,E_H)$ whose loss is $\alphafhrt$, and
moreover, one can efficiently (which means in polynomial time)
find an $H' = (K, E_{H'})$ whose loss is
$O(\frac{\log^2 k}{\log\log k})$.  They used these to give approximation
algorithms for several terminal-based problems, where the approximation
factor depended poly-logarithmically on the number of terminals $k$, and
not on $n$.  We note that they construct an arbitrary graph on $K$, and do
not attempt to directly obtain ``simple'' graphs; e.g., to get tree-based
flow-sparsifiers on $K$, they apply to $H'$ \Racke's method~\cite{R08}, and
increase the loss by an $O(\log k)$ factor.

In this paper, we simplify and unify some of these results: we show that
using the general framework of interchanging distance-preserving
mappings and capacity-preserving mappings from~\cite{R08}, which was
reinterpreted in an abstract setting by Andersen and Feige~\cite{AF09},
we obtain the following improvements over the results of~\cite{Moitra09,LM10}.
\procsonly{
\footnote{
Recently, it has come to our attention that, independent of and concurrent to our work,
Charikar, Leighton, Li, and Moitra, and independently Makarychev and Makarychev,
obtained results similar to the first two below,
as well as related lower bounds.}}

\begin{enumerate} \compactify
\item We show that using the $0$-extension results of~\cite{CKR04, FHRT03}
  in the framework of~\cite{R08,AF09} almost immediately gives us
  \emph{efficent} constructions of flow-sparsifiers with loss
  $\alphafhrt$. While the existential result of~\cite{LM10} also used
  the connection between $0$-extensions and flow-sparsifiers, the
  algorithmically-efficient version of the result was done \emph{ab initio},
  increasing the loss by another $O(\log k)$ factor. We use
  existing machinery, thereby simplifying the exposition somewhat, and
  avoiding the increased loss.
  See \lref[Theorem]{thm:gen-flowsp}.

\item We next use a randomized tree-embedding due to~\cite{GNR-reqcut},
  which is a variant of the so-called FRT tree-embedding~\cite{FRT03} where
  the expected stretch is reduced to $O(\log k)$ by requiring the
  non-contraction condition only for terminal pairs.  Using this refined
  embedding in the framework of~\cite{R08,AF09}, we obtain 
  in \lref[Theorem]{thm:cong-restate} efficient
  constructions of \emph{tree-based flow-sparsifiers} with loss $O(\log
  k)$.

\item We then turn to special families of graphs. For planar graphs, we
  give a new \zex algorithm that outputs a convex combination of
  $0$-extensions $f:V\to K$ (with $f(x)=x$ for all $x\in K$), such
  that all the corresponding \zex graphs $H_f = (K,E_f)$ (namely,
  $E_f=\{(f(u),f(v)):\ (u,v)\in E\}$) are \emph{planar graphs}, and its
  expected stretch $\max_{u,v \in V}  \nicefrac{\EX[d_{H_f}(f(u),f(v))
    ]}{d_G(u,v)} = O(1)$.  In particular, the planar graphs $H_f$
  produced are graph-theoretic minors of $G$. 
  These results are shown in \lref[Section]{sec:planar}.
  We remark that the known
  $0$-extension algorithms~\cite{CKR04,AFHKTT,LN05} do not ensure
  planarity of $H_f$.

  It follows that planar graphs admit a \emph{planar-based
    flow-sparsifier} (i.e., it is a convex combination of capacitated
  planar graphs on vertex-set $K$) with loss $O(1)$, and that we can
  find these efficiently.
  The fact that flow-sparsifiers with this
  loss \emph{exist} was shown by \cite{LM10}, but their sparsifiers are not planar-based.

  Moreover, the \zex algorithm itself can be viewed as a randomized
  version of Steiner point removal in metrics: previously, it was only
  known how to remove Steiner points from tree metrics with $O(1)$
  distortion~\cite{Gup01,CXKR06}. We believe this randomized procedure is of
  independent interest; e.g., combined with an embedding
  of~\cite{GNRS99}, this gives an alternate proof of the fact that the
  metric induced on the vertices of a single face of a planar graph can
  be embedded into a distribution over trees~\cite{LS09}.

\item The results for planar graphs are in fact much more general.
  Suppose $G$ is a $\beta_G$-decomposable graph (see definition in \lref[Section]{sec:notation}). Then we can efficiently output a
  distribution over graphs $H_f = (K,E_f)$ such that these are all
  minors of $G$, and the expected stretch is $$
\max_{u, v \in V} \frac{\EX[
    d_{H_f}(f(u),f(v)) ]}{d_G(u,v)} = O(\beta_G \log
  \beta_G).
$$ 
Now applying the same ideas of interchanging distance and
  capacity preservation, given any $G$ and $K$, 
  we construct in \lref[Corollary]{cor:minorsparsifiers}
  \emph{minor-based flow-sparsifiers} with loss $O(\beta_G \log
  \beta_G)$.

\item Finally, \lref[Section]{sec:lower-bounds} shows
  \fullonly{some} lower bounds on flow-sparsifiers: we show
  that flow-sparsifiers that are $0$-extensions of the original graph
  must have loss at least $\Omega(\sqrt{\log k})$ in the worst-case.
  For this class of possible flow-sparsifiers, this improves on the
  $\Omega(\log\log k)$ lower bound for sparsifiers proved
  in~\cite{LM10}. We also show that any flow-sparsifier that only uses
  edge capacities which are bounded from below by a constant, must
  suffer a loss of $\Omega(\sqrt{\log k}/\log\log k)$ in the worst-case.
\end{enumerate}

We can use these results to improve the approximation ratios 
of several application problems (see \lref[Section]{sec:applications}).
In many cases, constructions based on trees allow us to use better algorithms.
Our results are summarized in
\lref[Table]{tab:results}.  Note that apart from the two
linear-arrangement problems, our results smoothly approach the best
known results for the case $k=n$.

\begin{table}[htb]
\begin{center}
{\small
\newcolumntype{C}{>{\centering}X}
\renewcommand{\tabularxcolumn}[1]{m{#1}}
\begin{tabularx}{
\ifprocs
\else 0.95\textwidth
\fi
}{|@{~}>{\raggedright}m{3.8cm}@{~}|@{~}C@{~}|@{~}C@{~}|@{~}C@{~}|}
\hline\vgap{1pt}
& Previous Best Result & Our Result & Best Result when $k=n$\\ \hline\hline\vgap{1pt}
Flow-Sparsifiers (efficient) & $O(\frac{\log^2 k}{\log \log k})$ &
$O(\frac{\log k}{\log \log k})$ & --- \\ \vgap{1pt}\hline\vgap{1pt}
Tree-Based Flow-Sparsifiers & $O(\log n)^\dagger$, $O(\frac{\log^3 k}{\log\log k})$ &
$O(\log k)$ & $\Theta(\log n)$ \\ \vgap{1pt}\hline\vgap{1pt}
Minor-based Flow-Sparsifiers & --- &
$O(\beta_G \log \beta_G)$ & --- \\ \hline\hline\vgap{1pt}
Steiner Oblivious Routing & $\Otilde(\log^2 k)$ & $O(\log k)$ & $\Theta(\log n)$ \\ \hline\vgap{1pt}
$\ell$-Multicut & $\Otilde(\log^3 k)$ & $O(\log k)$ & $O(\log n)$ \\ \hline\vgap{1pt}
Steiner Minimum Linear Arrangement (SMLA)& $\Otilde(\log^{2.5} k)$ & $O(\log k
\log \log k)$ & $O(\sqrt{\log n} \log \log n)$ \\ \hline\vgap{1pt}
SMLA in planar graphs & $\Otilde(\log^{1.5} k)$ & $O(\log\log k)$ &  $O(\log
\log n)$\\ \hline\vgap{1pt}
Steiner Min-Cut Linear Arrangement & $\Otilde(\log^{4} k)$ & $O(\log^2
k)$ & $O(\log^{1.5} n)$
\\ \hline\vgap{1pt}
Steiner Graph Bisection & $O(\log n)^\dagger, O(\frac{\log^3 k}{\log \log k})$ & $O(\log k)$ & $O(\log n)$
\\ \vgap{1pt}\hline
\end{tabularx}
}
\caption{\label{tab:results} \small Summary of our results. Previous results
  marked with $\dagger$ from~\cite{R08}, all others
  from~\cite{Moitra09,LM10}.  }
\end{center}
\vspace*{-0.5cm}
\end{table}

Many of these applications further improve when the graph comes from a
minor-closed family (and hence has good $\beta$-decompositions), e.g.,
for the Steiner Minimum Linear Arrangement problem on planar graphs, we
can get an $O(\log \log k)$-approximation by using our minor-based
flow-sparsifiers to reduce the problem to planar instances on the $k$
terminals. Finally, \fullonly{in \lref[Section]{sec:direct-algos}}
\procsonly{in the full version} we show how to get better approximations
for the Steiner linear arrangement problems above using direct
LP/SDP approaches.

\ifprocs\else

\subsection{Concurrent Work}
\label{sec:concurrent_work}

Concurrently and independently from our work, 
Charikar, Leighton, Li, and Moitra~\cite{CLLM10} and independently Makarychev and Makarychev~\cite{MM10}
gave an efficient construction for $O(\log k/\log\log k)$-quality flow-sparsifiers. This is the same as our first result. Furthermore, Charikar et al.~\cite{CLLM10} give $O(\log k)$-quality tree-based flow-sparsifiers, which is the same as our second result.

Makarychev and Makarychev~\cite{MM10} also consider the case of graphs that exclude a fixed minor. They make the existential result of Leighton and Moitra~\cite{LM10} constructive and provide $O(1)$-quality flow-sparsifiers for these graphs. This is related to our third result. However, our construction has the additional advantage that the resulting flow-sparsifiers are guaranteed to be graph-theoretic minors of the original graph. This, for instance, results in improved approximation guarantees for Steiner Minimum Linear Arrangement for planar graphs.

For cut-sparsifiers, a weaker notion than flow-sparsifiers \cite{LM10}, 
lower bounds of $\Omega(\sqrt[4]{\log k/\log\log k})$ and $\Omega(\sqrt[4]{\log k})$,
were given by \cite{MM10} and~\cite{CLLM10}, respectively. 
(The former bound was improved to $\Omega(\sqrt{\log k}/\log\log k)$
in a later version.)
Makarychev and Makarychev~\cite{MM10} show an additional
lower bound of $\Omega(\sqrt{\log k/\log\log k})$ for flow-sparsifiers,
and also establish an interesting connection between flow- and
cut-sparsifiers and Lipschitz extendability of maps in Banach spaces. 
Charikar et al.~\cite{CLLM10} also exhibit a family of graphs 
for which the (best possible) quality of cut-sparsifiers 
with the restriction to $0$-extensions is asymptotically larger 
than without such restriction.

\subsection{Subsequent Work}

Subsequent to our work and using different techniques,
Chuzhoy~\cite{C12} shows that if the sparsifier $H$ is allowed to
contain a (relatively small) number of non-terminal vertices, it is
possible to construct O(1)-quality cut-sparsifiers of size $O(C^3)$ in
time $n^{O(1)}\cdot 2^C$, and $O(1)$-quality and flow-sparsifiers of size
$C^{O(\log\log C)}$ in time $n^{O(\log C)}\cdot 2^C$, where $C$ is an upper bound 
on the sum of capacities of all edges incident to any single terminal.
Andoni, Gupta and Krauthgamer \cite{AGK14} obtained a flow-sparsifier,
of quality $1+\eps$, which in effect is a tradeoff between quality and size,
for a restricted family that includes bipartite graphs.

Our results and techniques have proved useful in obtaining or simplifying
other results. Lee, Mendel, and Moharrami~\cite{LMM13} use our results
to show an approximate version of the Okamura-Seymour theorem for
node-capacitated graphs. Chekuri, Shepherd, and Weibel~\cite{CSW13} study a
problem similar to the Okamura-Seymour theorem, but with fewer restrictions
on the demands. More specifically, they consider an undirected planar graph
$G$ and a set of demand pairs such that at least one vertex of each of the
pairs lies in one of the outer $k$ layers of $G$. They show that if, for
any cut in $G$, the size of the cut is at least as large as the number of
demand pairs that have exactly one vertex on each side of the cut, 
then the demands are integrally routable in $G$
with congestion $c^k$ for some universal constant $c$. Their proof also uses
our results (unpublished note referenced in~\cite{CSW13}).

Chuzhoy, Makarychev, Vijayaraghavan, and Zhou~\cite{CMVZ12} study the edge-connectivity $k$-route cut
problem. In this problem an undirected edge-weighted graph, a set of
demands consisting of pairs of vertices, and a number $k$ are given. The
goal is to compute a minimum-weight subset of edges such that removing
these edges lets the edge-connectivity of every demand pair drop below
$k$.  They give a polynomial-time bicriteria approximation of this
problem which uses our algorithm for the $\ell$-Multicut problem as a
building block to handle large values of $k$.

Recently, Kamma, Krauthgamer, and Nguyen \cite{KKN14} showed how to
remove Steiner points from arbitrary graphical metrics
(following the results of \cite{Gup01,CXKR06} for tree metrics),
and obtain a single minor of the input graph that achieves a polylogarithmic stretch 
(distortion) for all terminal-terminal distances.
This result is incomparable to our result of 
$O(\beta_G \log \beta_G)$ expected stretch --- our bound on the stretch is better, but our guarantee is only for the expected stretch for any fixed pair of terminals.

\fi
\subsection{Notation}
\label{sec:notation}

Our graphs will have edge lengths or capacities; all edge lengths will
be denoted by $\ell: E \to \Rplus$, and edge costs/capacities will be
denoted by $c: E \to \Rplus$. When we refer to a graph $(G, \ell)$, we
mean a graph $G$ with edge lengths $\ell(\cdot)$; similarly $(H, c)$
denotes one with capacities $c(\cdot)$. When there is potential for
confusion, we will add subscripts (e.g., $c_H(\cdot)$ or
$\ell_G(\cdot)$) for disambiguation.  Given a graph $(G, \ell)$, the
shortest-path distances under the edge lengths $\ell$ is denoted by
$d_G: V \times V \to \Rplus$.

Given a graph $G = (V,E)$ and a subset of vertices $K \sse V$ designated
as \emph{terminals}, a \emph{retraction} is a map $f: V \to K$ such that
$f(x) = x$ for all $x \in K$.  For $(G,c)$ and terminals $K \sse V$, a
\emph{$K$-flow in $G$} is a multicommodity flow whose sources and sinks
lie in $K$.

\medskip
\noindent
\emph{Decomposition of Metrics.} 
Let $(X,d)$ be a metric space. 
A partition (i.e., a set of disjoint
``clusters'') $P$ of $X$ is called \emph{$\Delta$-bounded} if every
cluster $S\in P$ satisfies $\max_{u,v \in S}d(u,v) \leq \Delta$.  The
metric $(X,d)$ is called \emph{$\beta$-decomposable}
if for every $\Delta>0$ there is polynomial-time algorithm to sample
from a probability distribution $\mu$ over partitions of $X$, with the
following properties:
\begin{itemize}
\item[$\bullet$] \emph{Diameter bound:} Every partition $P\in\supp(\mu)$
  is $\Delta$-bounded.
\item[$\bullet$] \emph{Separation event:} For all $u,v\in X$, $\Pr_{P \in \mu}
  [\mbox{$\exists S\in P$ such that $u\in S$ but $v\notin S$}] \leq \beta\cdot d(u,v)/\Delta.$
\end{itemize}
$\beta$-decompositions of metrics have become standard tools with many applications; for more information see, e.g., \cite{LN05}.

When the metric arises as the shortest-path distances $d_G$ 
in a graph $G$ with nonnegative edge lengths $\ell$, 
we may assume that
each cluster $S$ in every partition $P$ in the support of $\mu$
induces a \emph{connected}
subgraph of $G$; if not, break such a cluster into its connected
components. The diameter bound and separation probabilities for edges
remain unchanged by this operation; indeed, the diameter bound is obvious,
and the separation probability for a non-adjacent pair $(u,v)$ 
(and similarly when $d_G(u,v)<\ell_G(u,v)$)
can be bounded by $\beta\cdot d_G(u,v)/\Delta$ by fixing a $u$-$v$ shortest path 
and noting that for $(u,v)$ to be separated, 
some shortest-path edge must be separated, and then applying the union bound.

We say that \emph{a graph} $G=(V,E)$ is $\beta$-decomposable if for
every assignment of nonnegative lengths $\ell$ to the edges, 
the resulting shortest-path
metric $d_{G}$ is $\beta$-decomposable.

\section{\texorpdfstring{$0$-Extensions}{0-Extensions}}
\label{sec:zextension}
In this section we provide a definition of 0-extension which is somewhat
different than the standard definition, and review some known results for
0-extensions. In \lref[Corollary]{thm:treesonk}, we also derive a variation of a
known result on tree embeddings, which will be applied in
\lref[Section]{sec:sparsifier-constructions}.

A \emph{$0$-extension} of graph $(G = (V,E), \ell_G)$ with terminals $K \sse V$
is usually defined as a retraction $f: V \to K$. We define a $0$-extension to
be a retraction $f: V \to K$ along with another graph $(H = (K, E_H), \ell_H)$;
here, the length function $\ell_H:E_H\to \R_+$ is defined as $\ell_H(x,y) =
d_G(x,y)$ for every edge $(x,y) \in E_H$. Note that this immediately implies
$d_H(x,y) \geq d_G(x,y)$ for all $x, y \in K$. Note also that $H_f$ defined in
\lref[Section]{sec:introduction} is a special case of $H$ in which
$E_H=\{(f(u),f(v)):(u,v)\in E\}$, whereas, in general, $H$ is allowed more
flexibility (e.g., $H$ can be a tree). This flexibility is precisely the reason
we are interested both in the retraction $f$ and in the graph $H$---we will
often want $H$ to be structurally simpler than $G$ (just like we want a
flow-sparsifier to be simpler than the original graph).

For a (randomized) algorithm $\A$ that takes as input $(G, \ell_G)$ and outputs
a (random) $0$-extension $(H, \ell_H)$, the \emph{stretch factor} of
algorithm $\A$ is the minimum $\alpha\ge 1$ such that
\[
\EX_H[\, d_H(f(x), f(y))\, ] \leq \alpha \, d_G(x,y) \qquad \text{for all }
x,y \in V.
\]
The following are well-known results for $0$-extension.
\begin{theorem}[\cite{FHRT03}]
  \label{thm:fhrt}
  There is an algorithm $\A_{FHRT}$ for $0$-extension with
  stretch factor $\alpha_{FHRT} = \alphafhrt$.
\end{theorem}
\begin{theorem}[\cite{CKR04}, see also~\cite{LN05}]
  \label{thm:leenaor}
  For graphs $G$ that are $\beta$-decomposable, there is an algorithm $\A_{CKR}$
  for $0$-extension with stretch factor $\alpha_{CKR} = O(\beta)$.
\end{theorem}
In particular, if the graph $G$ belongs to a non-trivial family of
graphs that is minor-closed, it follows from \cite{KPR93,FT03} that $\alpha = O(1)$.

\subsection{\texorpdfstring{$0$-Extension with Trees}{0-Extension with Trees}}

\ifprocs
\else
The following result is an extension of the tree-embedding theorem of
Fakcharoenphol et al.~\cite{FRT03}, where the difference is that the
following result ensures the non-contracting property~(a) only for
terminal-terminal pairs, but replaces the $O(\log n)$ by $O(\log k)$ in
the expected stretch between \emph{any} pair of nodes.
In what follows, a $c$-HST (abbreviation for Hierarchically Separated Tree) 
is a rooted tree with edge lengths, that satisfies the following for some $D>0$:
the distance between every leaf and its ancestor at level $j\ge0$ 
is exactly $D/c^j$.
(As usual, level means hop-distance from the root.)

\begin{theorem}[Tree embedding~\cite{GNR-reqcut}]
  \label{thm:logk-GNR}
  There is a randomized polynomial-time algorithm that takes as input a
  graph $G = (V,E)$ with terminals $K \sse V$ and outputs a (random)
  edge-weighted $2$-HST $T = (I \cup L, E_T)$ with internal nodes $I$
  and leaves $L$, and a map $f: V \to L$, such that
  \begin{OneLiners}
  \item[(a)] $d_T(f(x),f(y)) \geq d_G(x,y)$ for all $x, y \in K$ (with
    probability $1$),
  \item[(b)] $\EX_T[ d_T(f(x),f(y)) ] \leq O(\log k) \; d_G(x,y)$ for all
    $x, y \in V$, and
  \item[(c)] for each non-terminal $v \in V \setminus K$, either there
    exists a terminal $x_v$ sharing the leaf node with it (i.e., $f(v) =
    f(x_v)$), or another descendent of $f(v)$'s parent in $T$ contains a
    terminal $x_v$.
  \end{OneLiners}
\end{theorem}
\fi

\begin{corollary}[Tree \zex]
  \label{thm:treesonk}
  There is a randomized polynomial-time algorithm $\A_{GNR}$ for
  $0$-extension that has stretch factor $\alpha_{GNR} = O(\log k)$; furthermore, the
  graphs output by the algorithm are trees on the vertex set $K$.
\end{corollary}

\ifprocs
\else
\begin{proof}
  We need to give an algorithm that takes as
  input a graph $G = (V,E)$ with terminals $K \sse V$ and outputs a
  (random) edge-weighted tree $T = (K, E)$ and a retraction $f: V \to K$
  such that
  \begin{OneLiners}
  \item[\emph{(a')}] $d_T(x,y) \geq d_G(x,y)$ for all $x, y \in K$ (with
    probability $1$),
  \item[\emph{(b')}] $\EX_T[ d_T(f(x),f(y)) ] \leq O(\log k) \; d_G(x,y)$ for all
    $x, y \in V$.
  \end{OneLiners}
  
  We may assume that in $G$, all terminals are at non-zero distance
  from each other; otherwise, we can remove some terminals (from
  $K$, without changing $G$), apply the proof below, and add the
  terminals back in at the end.

  We start with sampling from the distribution of
  \lref[Theorem]{thm:logk-GNR} a random tree $T' = (I \cup L, E')$ and
  an associated map $f$.  We can take any leaf $l \in L$ whose
  pre-image set only contains non-terminals, remove the leaf, and remap
  all $v \in f^{-1}(l)$ to some other leaf that is a descendent of $l$'s
  parent node and also contains a terminal. (Such a leaf is guaranteed
  to exist by property~\emph{(c)} of \lref[Theorem]{thm:logk-GNR}.)
  While both the tree and the map change, we continue to call the
  modified tree $T'$ and the map $f$. We repeat this process until all
  leaves in the modified tree $T'$ contain at least one terminal.  
  Now property~\emph{(a)} implies
  (recall that in $G$, the distances between all terminals were nonzero)
  that each leaf contains at most one terminal.  Hence $f|_K$ is a $1$-$1$
  correspondence between the terminal set $K$ and the remaining leaves
  in the tree $T'$. Since the tree $T'$ is a $2$-HST, the distances in
  the tree between a remapped non-terminal and any other node in $T'$
  (apart from the one it was identified with) do not change.

  We can now remove all internal nodes in the modified version of $T'$
  (using, say,~\cite{Gup01}) to get a tree $T'' = (L, E'')$ on just the
  (erstwhile) leaves such that none of the $f(u)$-$f(v)$ distances are
  shrunk, and they are stretched by a factor of at most $8$.  The
  bijection between the set $L$ and terminals $K$ allows us to view the
  tree $T''$ as being on the node set $K$, and the map $f$ as being a
  retraction from $V \to K$.  Finally, shrinking the edges of the tree
  $T''$ only makes the expected stretch smaller, so we can reduce the
  length of any tree edge $e = (x,y)$ in $T''$ and set it equal to
  $d_G(x,y)$. Call this final tree $T$; it is immediate from
  properties~\emph{(a)} and~\emph{(b)} that this random $T$ and the
  associated retraction $f: V \to K$ satisfy properties~\emph{(a')}
  and~\emph{(b')} above, where the big-Oh term in property~(b') hides an
  extra stretch of $8$ due to this post-processing.
\end{proof}
\fi

As an aside, a weaker version of \lref[Corollary]{thm:treesonk} with
$O(\smash{\frac{\log^2 k}{\log \log k}})$ can be proved as follows.
First use \lref[Theorem]{thm:fhrt} to obtain a random $0$-extension $H$
from $G$ such that $\EX_H[ d_H(x,y) ] \leq \alphafhrt\; d_G(x,y)$ for all
$x,y \in K$. Then use the result of~\cite{FRT03} to get a random tree
$H' = (K, E_{H'})$ such that $\EX_{H'}[d_{H'}(x,y)] \leq O(\log k)\; d_{H}(x,y)$ for
all $x,y \in V(H)$. Combining these two results proves the weaker claim.

\section{Flow-Sparsifiers via \texorpdfstring{$0$-Extensions}{0-Extensions} }
\label{sec:using-trees}

In this section we first present the general framework of interchanging 
distance-preserving mappings and capacity-preserving mappings from~\cite{R08}, 
and its more abstract interpretation by Andersen and Feige~\cite{AF09},
and then discuss an algorithmically efficient implementations of it.
We then apply this framework, 
and ``transfer'' the results of \lref[Section]{sec:zextension},
which are aimed at preserving distances,
to results about preserving capacities,
which are essentially constructions of flow-sparsifiers.

Recall that given an edge-capacitated graph $(G,c)$ and a set $K \sse V$
of terminals, a \emph{flow-sparsifier with quality $\rho\ge1$} is another
capacitated graph $(H = (K, E_H), c_H)$ such that (a)~any feasible
$K$-flow in $G$ can be feasibly routed in $H$, and (b)~any feasible
$K$-flow in $H$ can be routed in $G$ with congestion $\rho$.

\subsection{Interchanging Distance and Capacity}
\label{sec:dist-cap}

We \fullonly{now} use the framework of R\"acke~\cite{R08}, as
interpreted by Andersen and Feige~\cite{AF09}. Given a graph $G =
(V,E)$, let $\P$ be a collection of multisets of $E$, which will
henceforth be called \emph{paths}. A mapping $M: E \to \P$ maps each
edge $e$ to a path $M(e)$ in $\P$. Such a map can be represented as a
matrix $\Mat$ in $\Z^{E \times E}$ where $\Mat_{e,e'}$ is the number
of times the edge $e'$ appears in the path (multiset) $M(e)$.  Given a
collection $\M$ of mappings (which we call the \emph{admissible}
mappings), a \emph{probabilistic mapping} is a probability distribution
over (or, convex combination of) admissible mappings; i.e., define $\lambda_M
\geq 0$ for each $M \in \M$ such that $\sum_{M \in \M} \lambda_M = 1$.

\paragraph{Distance Mappings.}
Given a graph $G = (V,E)$ with edge lengths $\ell: E \to \R_{> 0}$,
\begin{OneLiners}
\item the \emph{stretch} of an edge $e \in E$ under a mapping $M$ is $\sum_{e'}
  \Mat_{e,e'} \ell(e')/\ell(e)$.
\item the \emph{average stretch} of $e$ under a probabilistic mapping
  $\{\lambda_M\}$ is $\sum_M \lambda_M (\sum_{e'} \Mat_{e,e'}
  \ifprocs\else{\ell(e')}/{\ell(e)}\fi)$.
\item the \emph{stretch of a probabilistic mapping} is the maximum over all
  edges of their average stretch.
\end{OneLiners}

\paragraph{Capacity Mappings.}

Given a graph $G=(V,E)$ with edge capacities $c: E \to \R_{> 0}$,
  \begin{OneLiners}
  \item the \emph{load} of an edge $e' \in E$ under a mapping $M$ is
    $\sum_{e} \Mat_{e,e'} c(e)/c(e')$.
  \item the \emph{expected load} of $e'$ under a probabilistic mapping
    $\{\lambda_M\}$ is $\sum_M \lambda_M (\sum_{e} \Mat_{e,e'}
    \ifprocs\else{c(e)}/{c(e')}\fi)$.
  \item the \emph{congestion of a probabilistic mapping} is the maximum over
    all edges of their expected loads.
  \end{OneLiners}

\paragraph{The Transfer Theorem.}

Andersen and Feige~\cite{AF09} distilled ideas from
R\"acke~\cite{R08} to state:

\begin{theorem}[\protect{\cite[Theorem 6]{AF09}}]
  \label{thm:af-thm}
  Fix a graph $G=(V,E)$ and a collection $\M$ of admissible mappings.
  For every $\rho \geq 1$, the following are equivalent:
  \begin{enumerate}\compactify
    \item[1.] For every collection of edge lengths $\ell(\cdot)$, there is a
      probabilistic mapping with stretch at most $\rho$.
    \item[2.] For every collection of edge capacities $c(\cdot)$, there is a
      probabilistic mapping with congestion at most $\rho$.
  \end{enumerate}
\end{theorem}

Andersen and Feige~\cite{AF09} also outline how to make this result algorithmic:
if one can efficiently sample from the
probabilistic distance mapping with stretch $\rho$ (which is true for the
settings in this paper), one can efficiently sample from a probabilistic capacity
mapping with congestion $O(\rho)$ (and \emph{vice versa}). In fact, one
can obtain an explicit distribution on polynomially many admissible
mappings.
The techniques of  R\"{a}cke~\cite{R08} can also be used to obtain
this algorithmic version of the transfer theorem. Merely for completeness,
in the following we show how to derive the algorithmic result from a special
case of a theorem by Khandekar~\cite{K04}.
\begin{theorem}[\protect{\cite[Theorem 5.1.6]{K04}}]\label{thm:Khandekar}
Let $P\subseteq\R^d$ be a non-empty convex set for some $d$, 
and for each $e\in E$, let $f_e: P\to\R_{\ge0}$ be a non-negative
continuous convex function. Suppose we have an oracle that, given a vector
$x\in\R^E_{\ge0}$ with $\sum_{e\in E} x_e=1$ 
finds $\lambda\in P$ such that
$\sum_{e\in E} x_e f_e(\lambda) \le \rho$.
Then there exists an algorithm that given an error parameter $\omega\in (0,1)$
computes $\lambda\in P$ such that $\max_{e\in E} f_e(\lambda) \le e^\omega \rho$,
while making $O(\omega^{-2} m\log m)$ calls to the oracle and an
equal number of evaluations of $f_e(\cdot)$, where $m=|E|$.
\end{theorem}

This theorem can be used to show the following algorithmic version
of the transfer theorem.
\begin{corollary}\label{cor:af-thm-constructive}
Fix a graph $G=(V,E)$ and a collection $\mathcal{M}$ of admissible mappings. For
every $\rho\ge1$ and constant $\omega\in(0,1)$:

\begin{enumerate}[(a)]
\item 
Suppose that for every collection of edge lengths $\ell(\cdot)$ (edge capacities
$c(\cdot)$) there is an efficient algorithm to \textbf{compute} a probabilistic mapping
with stretch (congestion) at most $\rho$.
Then for every collection of
edge capacities $c(\cdot)$ (edge lengths $\ell(\cdot)$) there exists an efficient
algorithm to compute a probabilistic mapping with congestion (stretch) at
most $e^\omega \rho$.

\item 
Suppose that for every collection of edge lengths $\ell(\cdot)$ (edge capacities
$c(\cdot)$) there is an efficient algorithm to \textbf{sample} from a probabilistic
mapping with stretch (congestion) at most $\rho$.
Then for every
collection of edge capacities $c(\cdot)$ (edge lengths $\ell(\cdot)$) there exists
an efficient algorithm to compute a probabilistic mapping 
whose congestion (stretch) is, with high probability and in expectation,
at most $e^{2\omega} \rho+1$.
\end{enumerate}
\end{corollary}
\begin{proof}
We will show how to obtain a low-congestion probabilistic mapping if we
can, for every collection of edge lengths $\ell(\cdot)$, efficiently
compute (or sample from) a probabilistic mapping with low stretch. The
other direction, i.e., obtaining low stretch when we have a method to
obtain low-congestion probabilistic mappings, is symmetric.

\begin{enumerate}[(a)]
\item We define $f_{e'}(\lambda) := \sum_M \lambda_M (\sum_{e} \Mat_{e,e'}
c(e)/c(e'))$ to be the expected load of an edge $e'\in E$ under probabilistic
mapping $\{\lambda_M\}$ and we choose $P$ to be the set of all non-negative
$|\mathcal{M}|$-dimensional vectors $\lambda$ with $\sum_{M\in\mathcal{M}} \lambda_M =1$. Now
\lref[Theorem]{thm:Khandekar} immediately implies the claim if we can implement
the oracle efficiently.

Define edge lengths $\ell(e):= x_e/c(e)$. Due to our assumption, we can
efficiently find a probabilistic mapping $\{\lambda_M\}$ such that the
maximum average stretch, with respect to these edge lengths,
is at most $\rho$, i.e., such that
\[
\max_e \sum_M
\lambda_M \Big(\sum_{e'} \Mat_{e,e'} \frac{\ell(e')}{\ell(e)}\Big) \le \rho \enspace.
\]
Plugging in
$\ell(\cdot)$, we obtain
\[
\max_e \sum_M \lambda_M \Big(\sum_{e'}\Mat_{e,e'}
\frac{\ell(e')}{\ell(e)}\Big) = \max_e \frac{1}{x_e} \sum_M \Big(\lambda_M \sum_{e'}
x_{e'} \cdot \Mat_{e,e'} \frac{c(e)}{c(e')}\Big)\le\rho \enspace.
\]
Therefore, we can find
$\{\lambda_M\}$ such that, for every $e$, $\sum_M \lambda_M (\sum_{e'} x_{e'}
\cdot \Mat_{e,e'}c(e)/c(e')) \le \rho\cdot x_e $. Summing up over all $e$ gives
$\sum_e\sum_M \lambda_M (\sum_{e'} x_{e'} \Mat_{e,e'} c(e)/c(e')) \le \rho
\sum_e x_e = \rho$ and hence, by rearranging the sums,
\[
\sum_{e'}
x_{e'}\Big(\sum_M \lambda_M \Big(\sum_e \Mat_{e,e'} \frac{c(e)}{c(e')}\Big)\Big)=\sum_{e'} x_{e'}
f_{e'}(\lambda) \le \rho \enspace .
\]
This completes the implementation of the oracle.

\item Above we assumed that we can efficiently \emph{compute} an explicit distribution
on polynomially many admissible mappings that results in a probabilistic
mapping with low stretch. If we can only efficiently \emph{sample} from such a
distribution $\{\lambda_M\}$, we can still obtain a similar result. Let $C$ be an upper
bound on the worst load of any edge under any admissible mapping (e.g., the
maximum sum of all entries of an $M\in\mathcal{M}$ multiplied by the
largest ratio of capacities of two different edges). Then, for a
sufficiently large constant $\kappa$ we take $T=\ln (m C/\omega)\cdot \kappa/\omega$
independent samples from $\{\lambda_M\}$ and pick the sampled $M'\in\mathcal{M}$ that
minimizes $\sum_{e'} x_{e'} (\sum_e \mathbf{M'}_{e,e'} c(e)/c(e'))$. 
Our oracle then returns $\lambda'$ with $\lambda'_{M'}=1$ (and $\lambda'_{M''}=0$ for all
$M''\neq M'$).

For a single sample, the probability that $\sum_{e'} x_{e'} f_{e'}(\lambda') > e^\omega\rho$ is at most
$1/e^\omega$ due to Markov's inequality. The probability that this is the case for all $T$ independent samples is
at most $1/e^{\omega T}=(mC/\omega)^{-\kappa}$.
By taking a union bound over all
$O(\omega^{-2}m\log m)$ oracle calls we conclude that the probability that
any of them returns $\lambda'$ with $\sum_{e'} x_{e'} f_{e'}(\lambda') >
e^\omega\rho$ is bounded by $O((mC)^{2-\kappa})$. Therefore,
\lref[Theorem]{thm:Khandekar} guarantees that with high probability, namely
with probability at least $1- O((mC)^{2-\kappa})$, we obtain a
$\gamma \in P$ with $\max_e f_e(\gamma) \le e^{2\omega}\rho$.
With the remaining probability $O((mC)^{2-\kappa})$, 
$\max_e f_e(\gamma)$ may be much larger, 
but even in the worst case it will be bounded by $C$.
Therefore, by choosing $\kappa$
sufficiently large, the expectation of $\max_e f_e(\gamma)$ is bounded by $e^{2\omega}\rho+C\cdot O((mC)^{2-\kappa}) \le
e^{2\omega}\rho + 1$.
\end{enumerate}

\end{proof}

\subsection{Constructing Sparsifiers}\label{sec:sparsifier-constructions}
The following theorem gives the formal connection between $0$-extensions and 
flow sparsifiers.

\begin{theorem}
\label{thm:extensionstosparsifiers}
  Suppose there is a (randomized) algorithm $A$ that, 
  given a graph $G$ and edge lengths 
  $\ell_G: E(G)\rightarrow \mathbb{R}^+$,
  computes a \zex $((H,\ell_H),f)$ with stretch factor at most $\alpha$ such that $H$ is a graph
  from class $\mathcal H$.
 
  Then there is an algorithm that, given any capacity assignment
  $c_G:E(G)\rightarrow \mathbb{R}^+$, computes for the graph $(G,c_G)$ 
  an $O(\alpha)$-loss flow sparsifier 
  that is a convex combination of edge-capacitated graphs from class
  $\mathcal H$.
\end{theorem}

\begin{proof}
Suppose we have a \zex $(H,f)$, where $H=(K,E_H)$ and $f:V\rightarrow K$ is a
retraction. For every pair of terminals $u,v\in K$ we fix a canonical shortest
path $S_{u,v}^H$ between $u$ and $v$ in $H$ and a canonical shortest path
$S_{u,v}^G$ between $u$ and $v$ in $G$ (observe that for the important case
that $\mathcal H$ is the set of trees the paths in $H$ are unique). We define a
mapping $M_{H,f}:E(G)\rightarrow \mathcal P$ corresponding to \zex $(H,f)$ by
\begin{equation*} \label{eq:1}
M_{H,f}((x,y))=\biguplus_{(u,v)\in S_{f(x)f(y)}^H}S_{uv}^G\enspace.
\end{equation*}
In other words an edge $(x,y)$ is first mapped to $S^H_{f(x)f(y)}$ in $H$ and
then the edges $(u,v)$ on this path are mapped to path $S^G_{uv}$ in $G$.
Recall that $M_{H,f}((x,y))$ is a multi-set. In the corresponding matrix
representation, $\Mat_{e,e'}$ is the multiplicity of $e'$ in the set 
$\uplus_{(u,v)\in S_{f(x)f(y)}^H}S_{uv}^G$.

For a graph class $\mathcal H$ (for example the set of trees) we define the set
of \emph{admissible} mappings by $\{M_{H,f} \mid H\in\mathcal H\}$. Note
that in $M_{H,f}$ an edge $(x,y)\in E(G)$ is mapped to a path of length
$d_H(f(x),f(y))$. This means the stretch of the edge in the mapping is the same
as the stretch of an edge in the definition of $0$-extensions. Therefore, the
existence of a probability distribution over $0$-extensions with (expected) stretch
$\alpha$ gives rise to a probability distribution over admissible mappings with
(expected) stretch $\alpha$.

Applying the constructive version of the Transfer Theorem gives that for any
assignment $c_G:E(G)\rightarrow\mathbb{R}^+$ of edge capacities to edges in
$G$, we can compute a probability distribution over admissible mappings with
congestion at most $O(\alpha)$. In the following we show that we can interpret
this probability distribution as a flow-sparsifier. 

With every mapping $M_{H,f}$ we associate the graph $H$ with the following
edge capacities
\begin{equation*}
c_{H,f}(e) = \sum_{(u,v)\in E(G):e\in S_{f(u),f(v)}^H}c_G((u,v))\enspace.
\end{equation*}
This means the capacity of an edge $e\in E(H)$ is the total capacity of all graph
edges $(u,v)\in G$, for which the canonical path between $u$ and $v$ in $H$ contains
$e$. The flow sparsifier $F$ is now the convex combination $\{\lambda_{H,f}\}$ over
graphs $(H,c_{H,f})$. To see that $F$ has quality $O(\alpha)$ 
we prove two facts:
\begin{enumerate}[(a)]
\item any $K$-flow that can be feasibly routed in $G$, can also be feasibly
  routed in $F$; and 
\item any $K$-flow that can be feasibly routed in $F$, can be routed with
  congestion $O(\alpha)$ in $G$. 
\end{enumerate}
Proving these facts is essentially a matter of unraveling the definitions. 
For~(a), the definition of edge capacities $c_{H,f}$ ensures that $(H,c_{H,f})$ can
feasibly route all edges of $G$ concurrently. Hence, it can also route any
$K$-flow that is feasible in $G$. Since, this is true for any graph
$(H,c_{H,f})$ it also holds for the convex combination $F$.

To prove~(b), we want to route edges of $F$ in $G$. As $F$ is a convex
combination this means we want to concurrently route all graphs $(H,c_{H,f})$,
where the capacities are scaled down by the convex multiplier $\lambda_{H,f}$. 
We simply route an edge $(u,v)\in H$ along the canonical path $S^G_{uv}$. 
This results in the following load on an edge $e'\in E(G)$:
\begin{equation*}
\frac{1}{c(e')}\sum_{H,f}\lambda_{H,f}\sum_{e_H=(u,v)\in E(H):e'\in S_{u,v}^G}c_{H,f}(e_H)\enspace.
\end{equation*}
Plugging in the definition for the edge capacities $c_{H,f}$ and changing the
order of summation gives
that this is equal to 
\begin{equation*}
\begin{split}
&\frac{1}{c(e')}\sum_{H,f}\lambda_{H,f}\sum_{e_H = (u,v) \in E(H):
      e' \in S_{uv}^G}
    \sum_{(x,y) \in E(G): e_H \in S_{f(x),f(y)}^H} c(xy) \\
    = & \frac{1}{c(e')}\sum_{H,f}\lambda_{H,f}\sum_{(x,y) \in E} c(xy)
    \cdot (\text{multiplicity of $e'$ in } \uplus_{(u,v) \in
      S^H_{f(x),f(y)}} S_{uv}^G)\enspace.
\end{split}
\end{equation*}
However, this is exactly the \emph{expected load} for $e'$ under the notion of
admissible maps defined in~(\ref{eq:1}); hence this is bounded by the
congestion (the maximum expected load over all edges), which is at most
$O(\alpha)$. This proves condition~(b) above, that the congestion to route any
$K$-flow in the convex combination $F$ in the graph $G$ is at most
$O(\alpha)$. \procsonly{\qed}
\end{proof}

Combining \lref[Theorem]{thm:extensionstosparsifiers} with 
\lref[Corollary]{thm:treesonk} gives the following.

\begin{theorem}[Tree-based Flow-Sparsifiers]
\label{thm:cong-restate}
There is a randomized polynomial-time algorithm that,
given a graph $G$ and terminals $K$, 
outputs a flow-sparsifier $H$ which is a convex combination of trees
and has loss $O(\log k)$.
\end{theorem}

Combining \lref[Theorem]{thm:extensionstosparsifiers} with \lref[Theorem]{thm:fhrt} gives the following.
\begin{theorem}[Flow-Sparsifiers]
  \label{thm:gen-flowsp}
  There is a randomized polynomial-time algorithm that,
  given a graph $G$ with terminals $K$, 
  outputs a flow-sparsifier $H$ with loss $\alphafhrt$.
\end{theorem}

The same idea using \zex results for $\beta$-decomposable graphs
(\lref[Theorem]{thm:leenaor}) gives us the following.
\begin{theorem}[Flow-Sparsifiers for Minor-Closed Families]
  \label{thm:minor-flowsp}
  There is a randomized polynomial-time algorithm that,
  given a $\beta$-decomposable graph $G$ with terminals $K$, 
  constructs a flow-sparsifier with loss $O(\beta)$.
\end{theorem}
Note that the decomposability holds if $G$ belongs to a non-trivial
minor-closed-family $\mathcal{G}$ (e.g., if $G$ is planar).  However, 
\lref[Theorem]{thm:minor-flowsp} does not claim that the flow-sparsifier
for $G$ also belongs to the family $\mathcal{G}$; this is the
question we resolve in the next section.

\newcommand{\B}{\mathcal{B}}
\newcommand{\newbeta}{\smash{\hat{\beta}}}
\newcommand{\partialP}{\smash{\hat}{P}}

\section{\texorpdfstring{Connected $0$-Extensions and Minor-Based 
    Flow-Sparsifiers}{Connected 0-Extensions and Minor-Based Flow-Sparsifiers}}
\label{sec:planar}
The results in this section apply to $\beta$-decomposable graphs. A prominent example of such graphs are planar graphs, which (along with every family of graphs excluding a fixed minor) are $O(1)$-decomposable \cite{KPR93,FT03}. 
Thus, \lref[Theorem]{thm:connectedzextdec}, \lref[Corollary]{cor:minorsparsifiers} and \lref[Theorem]{thm:steinerremoval} below all apply to planar graphs (and more generally to excluded-minor graphs) with $\beta=O(1)$. We now state our results for $\beta$-decomposable graphs in general. In \lref[Section]{sec:term-decomp} we define a related notion called
\emph{terminal-decomposability}, and show analogous results for 
$\newbeta$-terminal-decomposable graphs. 

In what follows we use the definition of $0$-extension from
\lref[Section]{sec:zextension} with $H=H_{f}$, i.e.,
$E_H=\{(f(u),f(v)):(u,v)\in E\}$, hence the \zex is completely defined
by the retraction $f$. We say that a $0$-extension $f$ is {\em connected} if for every $x$, $f^{-1}(x)$ induces a connected component in $G$. 
Our main result shows that we get connected $0$-extensions with stretch $O(\beta\log\beta)$ for $\beta$-decomposable metrics. 

\begin{theorem}[Connected $0$-Extension]
  \label{thm:connectedzextdec}
  There is a randomized polynomial-time algorithm that, given $(G=(V,E), \ell_G)$ with terminals $K$ such that $d_G$ is
  $\beta$-decomposable, produces a connected $0$-extension $f:V\to K$ such
  that for all $u,v\in V$, we have
  $$\EX[d_{H}(f(u),f(v))] \leq O(\beta\log \beta)\cdot d_{G}(u,v).$$
\end{theorem}

\noindent
Note that if $f$ is a connected $0$-extension, the graph $H_f$ is a
minor of $G$. Applying \lref[Theorem]{thm:af-thm} to interchange the
distance preservation with capacity preservation, we get the following
analogue of \lref[Theorem]{thm:cong-restate}.
\begin{corollary}[Minor-Based Flow-Sparsifiers]
  \label{cor:minorsparsifiers}
  For every $\beta$-decomposable graph $G=(V,E)$ with edge capacities
  $c_{G}$ and a subset $K\subset V$ of $k$ terminals, there is a
  minor-based flow-sparsifier with quality $O(\beta \log \beta)$ .
  Moreover, a minor-based flow-sparsifier for $G,c_{G},K$ can be computed
  efficiently in randomized polynomial-time.
\end{corollary}

Since planar graphs are $O(1)$-decomposable and since their minors are planar,
by \lref[Corollary]{cor:minorsparsifiers} they have an efficiently constructable planar-based flow-sparsifier with quality $O(1)$. By \lref[Theorem]{thm:connectedzextdec}, they always have a connected $0$-extension with stretch at most $O(1)$. An interesting consequence of the latter result is that given any planar graph $(G,\ell_G)$, and a set $K$ of terminals, we can ``remove'' the non-terminals and get a related planar graph on $K$ while
preserving inter-terminal distances in expectation. 
Moreover, this extends to every family of graphs excluding a fixed
minor.
These results generalize a
result from~\cite{Gup01} showing a similar result for trees.\footnote{One difference from the result in~\cite{Gup01} is the following: that result deterministically produced a single tree after removing the non-terminals, and
hence the distances were preserved deterministically, and not just in expectation. Getting such a result for planar graphs remains an open problem.}

\begin{theorem}[Steiner Points Removal]
  \label{thm:steinerremoval}
  There is a randomized polynomial-time algorithm that, given $(G=(V,E),
  \ell_G)$ and $K$ such that $d_G$ is $\beta$-decomposable, outputs
  minors $H = (K, E_H)$ of $G$ such that $1 \leq \frac{
    \EX[d_H(x,y)]}{d_G(x,y)} \leq O(\beta \log \beta)$ for all $x,y \in
  K$.
\end{theorem}

\noindent
Note that, since general graphs are only $\Theta(\log n)$-decomposable,
these results only give us an $O(\log n \log\log
n)$-approximation for connected $0$-extension on arbitrary graphs (or an
$O(\log^2 k \log\log k)$-approximation using results of
\lref[Section]{sec:term-decomp}). We can improve that to $O(\log k)$;
\ifprocs\else the details are in
\lref[Section]{sec:ckr-conn}.  \fi

\begin{theorem}[Connected CKR]
  \label{thm:connectedzextgen}
  There is a randomized polynomial-time algorithm that on input
  $(G=(V,E), \ell_G)$ and $K$, produces a connected $0$-extension $f$
  with $\EX[d_{H}(f(u),f(v))] \leq O(\log k)\cdot d_{G}(u,v)$ for all
  $u,v\in V$.
\end{theorem}

Using the semi-metric relaxation for $0$-extension, we get a connected
$0$-extension whose cost is at most $O(\log k)$ times the optimal
(possibly disconnected) $0$-extension. To our knowledge, this is the
first approximation algorithm for connected $0$-extension, and in fact
shows that the gap between the optimum connected $0$-extension and the
optimum $0$-extension is bounded by $O(\log k)$. The same is true with
an $O(1)$ bound for planar graphs. We remark that the connected
$0$-extension problem is a special case of the connected metric labeling
problem, which has recently received attention in the vision
community~\cite{VicenteKR08,NowozinL09}.

\subsection{The Algorithm for Decomposable Metrics}

We now give the algorithm behind \lref[Theorem]{thm:connectedzextdec}.
Assume that edge lengths $\ell_G$ are integral and scaled such that the
shortest edge is of length 1. Let the diameter of the metric be at most
$2^{\delta}$. For each vertex $v \in V$, define $A_v = \min_{x \in K}
d_G(v,x)$ to be the distance to the closest terminal. The algorithm
maintains a partial mapping $f$ at each point in time---some of the
$f(v)$'s may be undefined (denoted by $f(v) = \bot$) during the run, but
$f$ is a well-defined \zex when the algorithm terminates.  We say a
vertex $v \in V$ is \emph{mapped} if $f(v) \neq \bot$. The algorithm
appears as Algorithm~1.

\begin{algorithm}
  \caption{Algorithm for Connected \zex}
  \begin{algorithmic}[1]
   \STATE {\bf input:} $(G, \ell_G), K$.
   \STATE \textbf{let} $i \gets 0$, $f(x) = x$ for all $x \in K$, $f(v) = \bot$
   for all $v \in V \setminus K$.

   \WHILE{there is a $v$ such that $f(v) = \bot$}

   \STATE \textbf{let} $i \gets i+1$, $r_i \gets 2^i$

   \STATE sample a $\beta$-decomposition of $d_G$ with diameter
   bound $r_{i}$ to get a partition $P$

   \FORALL{clusters $C_s$ in the partition $P$ that contains both mapped
     and unmapped vertices}

   \STATE delete all vertices $u$ in $C_s$ with $f(u) \neq \bot$

   \FOR{each connected component $C$ from $C_s$}

   \STATE choose a vertex $w_C \in C_s$ that was deleted and had an
   edge to $C$ \label{line:wnonsteiner}

   \STATE reset $f(u) = f(w_C)$ for all $u \in C$.

   \ENDFOR

   \ENDFOR

   \ENDWHILE

  \end{algorithmic}
\end{algorithm}

\procsonly{\vspace{-0.1in}}
\noindent
We can assume that in round $\delta = \log \diam(G)$, the partitioning
algorithm returns a single cluster, in which case all vertices are
mapped and the algorithm terminates. Let $f_{i}$ be the mapping at the
end of iteration $i$. For $x \in K$, let $V_i^x$ denote $f_i^{-1}(x)$,
the set of nodes mapped to $x$. The following claim follows inductively:
\begin{lemma}
  For every \fullonly{iteration}  $i$ and $x \in K$, the set $V_i^x$ induces a connected
  component in $G$.
\end{lemma}
\begin{proof}
  We prove the claim inductively. For $i=0$, there is nothing to prove
  since $V_i^x = \{x\}$.  Suppose that in iteration $i$, we map vertex
  $u$ to $x$ so that $u \in V_i^x$. Thus for some component $C$
  containing $u$, the mapped neighbor $w_C$ chosen by the algorithm was
  in $V_{i-1}^x$. Since we map all of $C$ to $x$, there is a path
  connecting $v$ to $w_C$ in $V_i^x$. Inductively, $w_C$ is connected to
  $x$ in $V_{i-1}^x \subseteq V_i^x$, and the claim follows.
\procsonly{\qed}\end{proof}

The following lemma will be useful in the analysis of the stretch; it
says that any node mapped in iteration $i$ is mapped to a terminal at
distance $O(2^i)$.
\begin{lemma}
  \label{lem:smallclusters}
  For every iteration $i$ and $x \in K$, and every $u \in V_i^x$, $d_G(x,u) \leq 2 r_i$.
\end{lemma}
\begin{proof}
  The proof is inductive. For $i=0$, the claim is immediate.  Suppose
  that in iteration $i$, we map vertex $u$ to $x$ so that $u \in V_i^x$.
  Thus for some component $C$ containing $u$, the mapped neighbor $w_C$
  chosen by the algorithm was in $V_{i-1}^x$. Moreover, $u$ and $w_C$
  were in the same cluster in the decomposition so that $d(u,w_C) \leq
  r_i$. Inductively, $d(w_C,x) \leq 2r_{i-1}$ and the claim follows by
  triangle inequality.
\procsonly{\qed}\end{proof}

In the remainder of the section, we bound the stretch of the $0$-extension;
for every edge $e=(u,v)$ of $G$, we show that
$$\EX[d_G(f(u),f(v)] \leq O(\beta \log \beta)\; d_G(u,v).$$
Note that for $e=(u,v)$, $d_G((f(u),f(v))=d_H((f(u),f(v))$. Therefore it is sufficient
to prove the claim for $d_G$. The analogous claim for
non-adjacent pairs will follow by triangle inequality, but here with
$d_H$. We say that the edge $e=(u,v)$ is \emph{settled in round $j$} if
the later of its endpoints gets mapped in this round; $e$ is
\emph{untouched after round $j$} if both $u$ and $v$ are unmapped at the end of round
$j$. Let $d_G(u,K) \leq d_G(v,K)$ and let $A_e$ denote the distance
$d_G(u,K)$. Let $j_e:=\lfloor\log(A_e)\rfloor-1$.

\begin{lemma} \label{lem:settled}
  For edge $e = (u,v)$,
  \begin{OneLiners}
    \item[(a)] edge $e$ is untouched after round $j_e-1$,
    \item[(b)] if edge $e$ is settled in round $j$ then $d_G(f(u),f(v)) =
      O(2^j+d_G(u,v))$.
  \end{OneLiners}
\end{lemma}

\begin{proof}
  For~(a), if one of the end points of $e$ is mapped before round $j_e$,
  then $2\cdot 2^{j_e} \leq A_e = d_G(e,K)$, which
  contradicts \lref[Lemma]{lem:smallclusters}. For~(b), both $d_G(u,f(u)),
  d_G(v,f(v)) \leq 2^{j+1}$ by \lref[Lemma]{lem:smallclusters}; the
  triangle inequality completes the proof.
\procsonly{\qed}\end{proof}

Let $\B_j$ denote the ``bad'' event that the edge is settled in round
$j$ and that both end-points are mapped to different terminals. Let
$z:=\max\{A_e,d_G(u,v)\}$. We want to use 
$$\EX[d(f(u),f(v))] = \sum_{j}\Pr[\B_j]\cdot \EX[d(f(u),f(v))\mid \B_j].$$

\begin{claim}
  \label{cla:probxj}
  {\rm$\Pr[\B_j]\le\min\{4\beta\frac{z}{2^j},1\}\cdot5\beta\frac{d_G(u,v)}{2^j}$}.
\end{claim}

\begin{proof}
  Recall that an edge is untouched after round $j'$ if neither of
  its endpoints is mapped at the end of this round. For this to happen,
  $u$ must be separated from its closest terminal in the clustering in
  round $j'$, which happens with probability at most
  $\min\{\beta\frac{A_e}{2^{j'}},1\}$. Also recall that the probability
  that an edge $e=(u,v)$ is cut in a round $j'$ is at most
  $\beta\frac{d_G(u,v)}{2^{j'}}$.  Let $i$ denote the round in which the
  edge is first touched. We upper bound the probability of the event
  $\B_j$ separately depending on how $i$ and $j$ compare. Note that for
  $j \leq 2$, the right hand side is at least $1$ so the claim holds
  trivially.
  \begin{itemize}
  \item[$\bullet$] $i\le j-2$. For $\B_j$ to occur, the edge $e$ must be cut in
    round $j-2$ and $j-1$, as otherwise it would already be settled in
    one of these rounds. The probability of this is at most
    $\min\{\beta\frac{d_G(u,v)}{2^{j-2}},1\}\cdot\beta\frac{d_G(u,v)}{2^{j-1}}
    \le \min\{4\beta\frac{z}{2^{j}},1\}\cdot2\beta\frac{d_G(u,v)}{2^j}$.
  \item[$\bullet$] $i=j-1$. For $\B_j$ to occur, the edge $e$ must be cut in round
    $j-1$ and must be untouched after round $j-2$. The probability of this
    is at most
    $\min\{\beta\frac{A_e}{2^{j-2}},1\}\cdot\beta\frac{d_G(u,v)}{2^{j-1}}
    \le \min\{4\beta\frac{z}{2^{j}},1\}\cdot2\beta\frac{d_G(u,v)}{2^j}$.
  \item[$\bullet$] $i=j$. For $\B_j$ to occur, $e$ must be cut in round $j$ and
    must be untouched after round $j-1$. The probability of this is at most
    $\min\{\beta\frac{A_e}{2^{j-1}},1\}\cdot\beta\frac{d_G(u,v)}{2^{j}}
    \le \min\{4\beta\frac{z}{2^{j}},1\}\cdot\beta\frac{d_G(u,v)}{2^j}$.
  \end{itemize}
  Since $\Pr[\B_j] = \Pr[\B_j \wedge (i\leq j-2)] + \Pr[\B_j \wedge
  (i=j-1)] +\Pr[\B_j \wedge (i=j)]$, the claim follows.
\procsonly{\qed}\end{proof}

\noindent
\lref[Lemma]{lem:settled}(b) implies that if the edge is settled before
round $j_d:=\lfloor\log(d_G(u,v))\rfloor$, the conditional expectation
$\EX[d_G(f(u),f(v))\mid \B_j]$ is $O(d_G(u,v))$.  Moreover the
edge $e$ cannot be settled before round $j_e =\lfloor\log(A_e)\rfloor-1$
by \lref[Lemma]{lem:settled}(a). Let $j_m:=\max\{j_d,j_e\}$. It therefore
suffices to to show that
\begin{equation*}
\sum_{j\ge j_m}\Pr[\B_j]\cdot O(2^j)\le O(\beta\log\beta)\;d_G(u,v)\enspace.
\end{equation*}
Plugging in the upper bound for $\Pr[\B_j]$ into the left hand side, we get
\begin{equation*}
\begin{split}
\ts \sum_{j\ge j_m}\Pr[\B_j]\cdot O(2^j)
&\ts \le
\sum_{j\ge
  j_m}\min\{{\textstyle4\beta\frac{z}{2^j}},1\}\cdot{\textstyle5\beta\frac{d_G(u,v)}{2^j}}
\cdot O(2^j)\\
&\ts \le\sum_{j\ge
  j_m}\min\{{\textstyle4\beta\frac{z}{2^j}},1\}\cdot\beta
\cdot O(d_G(u,v)) \quad \le O(\beta\log\beta)\;d_G(u,v)\enspace.
\end{split}
\end{equation*}
In the last step, we used that $z = \max\{A_e, d_G(u,v)\} \leq
\max\{2^{j_e + 2}, 2^{j_d+1}\} \le 2^{j_m+2}$, so the first $O(\log
\beta)$ terms contribute $O(\beta\,d_G(u,v))$, while the remaining terms
form a geometric series and sum to  $O(d_G(u,v))$.  This
completes the proof of \lref[Theorem]{thm:connectedzextdec}.

\subsection{Terminal Decompositions}
\label{sec:term-decomp}

The general theorem for connected \zex{s} gives a guarantee in terms of its
decomposition parameter $\beta$, and in general this quantity may depend
on $n$. This seems wasteful, since we decompose the entire metric while
we mostly care about separating the terminals. 

To this end, we define \emph{terminal decompositions} (the reader might
find it useful to contrast it with definition of decompositions in
\lref[Section]{sec:notation}). A \emph{partial partition} of a set $X$
is a collection of disjoint subsets (called ``clusters'' of $X$). A
metric $(X,d)$ with terminals $K$ is called \emph{
  $\newbeta$-terminal-decomposable} if for every $\Delta>0$ there is
probability distribution $\mu$ over partial partitions of $X$, with the
following properties:
\begin{OneLiners}
\item \emph{Diameter bound:} Every partial partition
  $\widehat{P}\in\supp(\mu)$ is connected and $\Delta$-bounded.
\item \emph{Separation event:} For all $u,v\in X$, $
  \Pr_{\widehat{P}\in \mu} [\mbox{$\exists S\in \widehat{P}$ such that
    $u\in S$ but $v\notin S$}]  \leq \newbeta\cdot d(u,v)/\Delta.
  $
\item \emph{Terminal partition:} For all $x\in K$, every partial
  partition $\widehat{P}\in\supp(\mu)$ has a cluster containing $x$.
\item \emph{Terminal-centered clusters:} For every partial partition
  $\widehat{P}\in\supp(\mu)$, every cluster $S\in \widehat{P}$ contains
  a terminal.
\end{OneLiners}
A graph $G=(V,E)$ with terminals $K$ is $\newbeta$-terminal-decomposable if
for every nonnegative lengths $\ell_{G}$ assigned to its edges, the
resulting shortest-path metric $d_{G}$ with terminals $K$ is
$\newbeta$-terminal-decomposable.  Throughout, we assume that there is a
polynomial time algorithm that, given the metric, terminals and $\Delta$
as input, samples a partial partition $\widehat{P}\in\mu$.
Note that if $K=V$, the above definitions coincide with the definitions
of $\beta$-decomposable metrics and graphs.

Our main theorem for terminal decomposable metrics is the following:
\begin{theorem}
  \label{thm:connectedzextterminal}
  Given $(G=(V,E), \ell_G)$, suppose $d_G$ is
  $\newbeta$-terminal-decomposable with respect to terminals $K$. There
  is a randomized polynomial-time algorithm that produces a connected
  $0$-extension $f:V\to K$ such that for all $u,v\in V$, we have
  $\EX[d_{G}(f(u),f(v))] \leq O(\newbeta^2\log \newbeta)\cdot d_{G}(u,v).$
\end{theorem}
This theorem is interesting when $\newbeta$ is much less than $\beta$,
the decomposability of the metric itself. E.g., one can alter the CKR
decomposition scheme to get $\newbeta(k,n) = O(\log k)$, while $\beta =
O(\log n)$.

\procsonly{\vspace{-0.1in}}
\subsubsection{The Modified Algorithm.}

Algorithm~2 for the terminal-decomposable case is very similar to
Algorithm~1: the main difference is that in each iteration we only
obtain a partial partition of the vertices, we map only the nodes that
lie in clusters of this partial partition.

A few words about the algorithm: recall that a partial partition returns
a set of connected diameter-bounded clusters such that each cluster
contains at least one terminal, and each terminal is in exactly one
cluster--- we use $V^x$ to denote the cluster containing $x \in K$.
(Hence either $V^x = V^y$ or $V^x \cap V^y = \emptyset$.) Now when we
delete all the vertices in some cluster $V^x$ that are already mapped,
this includes the terminal $x$---and hence there is at least one
candidate for $w_C$ in \lref[Line]{line:wnonsteiner}.  Eventually, there will be
only one cluster, in which case all vertices are mapped and the
algorithm terminates.

\begin{algorithm}
  \caption{Algorithm for Connected \zex: the terminal-decomposable case}
  \begin{algorithmic}[1]
   \STATE {\bf input:} $(G, \ell_G), K$.
   \STATE \textbf{let} $i \gets 0$, $f(x) = x$ for all $x \in K$, $f(v) = \bot$
   for all $v \in V \setminus K$.

   \WHILE{there is a $v$ such that $f(v) = \bot$}

   \STATE \textbf{let} $i \gets i+1$, $r_i \gets 2^i$

   \STATE find a $\newbeta$-terminal-decomposition of $d_G$ with diameter
   bound $r_{i}$; let $V^x$ be the cluster containing terminal $x$.

   \FORALL{clusters $V^x$ in the partial partition}

   \STATE delete all vertices $u$ in $V^x$ with $f(u) \neq \bot$
   \FOR{each connected component $C$ from $V^x$ thus formed}

   \STATE choose a vertex $w_C \in V^x$ that was deleted and had a
   neighbor in $C$ \label{line:wmodified}

   \STATE reset $f(u) = f(w_C)$ for all $u \in C$.

   \ENDFOR

   \ENDFOR

   \ENDWHILE

  \end{algorithmic}
\end{algorithm}

\procsonly{\vspace{-0.1in}}
\noindent
The analysis for \lref[Theorem]{thm:connectedzextterminal} is almost the
same as for \lref[Theorem]{thm:connectedzextdec}; the only difference is
that \lref[Claim]{cla:probxj} is replaced by the following weaker claim
which immediately gives the $O(\newbeta^2 \log
\newbeta)$ bound.
\begin{claim}
  \label{cla:probxj-steiner}
  {\rm$\Pr[\B_j]\le\min\{8\newbeta\frac{z}{2^j},1\}\cdot23\newbeta^2\frac{d(u,v)}{2^j}$}.
\end{claim}

\ifprocs
\else
\begin{proof}
  Recall that an edge is \emph{untouched} after round $j'$ if neither of
  its endpoints is mapped at the end of this round. For this to happen,
  $u$ must be separated from it's closest terminal in the clustering in
  round $j'$, which happens with probability at most
  $\min\{\newbeta\frac{A_e}{2^{j'}},1\}$. Also recall that the probability
  that an edge $e=(u,v)$ is cut in a round $j'$ is at most
  $\newbeta\frac{d(u,v)}{2^{j'}}$.  Let $i$ denote the round in which the
  edge is first touched. We upper bound the probability of the event
  $\B_j$ separately depending on how $i$ and $j$ compare. Note that for
  $j \leq 3$, the right hand side is at least $1$ so the claim holds
  trivially.
  \begin{itemize}
  \item $i\le j-3$. For $\B_j$ to occur, it must happen that the edge is
    cut in round $i$ and it is either untouched or cut in rounds $j-1$ and
    $j-2$. The probability for this to happen is at most
$\min\{\newbeta\frac{d(u,v)}{2^i},1\}\cdot\min\{\newbeta(\frac{A_e}{2^{j-2}}+\frac{d(u,v)}{2^{j-2}}),1\}\cdot\newbeta(\frac{A_e}{2^{j-1}}+\frac{d(u,v)}{2^{j-1}})
\le\min\{\frac{d(u,v)}{2^i},1\}\min\{8\newbeta\frac{z}{2^{j}},1\}\cdot4\newbeta^2\frac{z}{2^{j}}$.
If $d(u,v)\ge A_e$ this is at most
$\min\{8\newbeta\frac{z}{2^{j}},1\}\cdot16\newbeta^2\frac{d(u,v)}{2^{j}}$ as
$z=d(u,v)$. Otherwise, observe that $i \ge j_e$ as the edge cannot be touched
before. Hence $2^i \ge A_e/4$, and plugging this in gives a bound of
$\min\{8\newbeta\frac{z}{2^{j}},1\}\cdot16\newbeta^2\frac{d(u,v)}{2^{j}}$, as well.

  \item $i= j-2$. For $\B_j$ to occur, the edge $e$ must be cut in
    round $j-2$ and it must be cut or untouched in round $j-1$, as otherwise it would already be settled in
    one of these rounds. The probability of this is at most
    $\newbeta\frac{d(u,v)}{2^{j-2}}\cdot\min\{\newbeta(\frac{d(u,v)}{2^{j-1}}+\frac{A_e}{2^{j-1}}),1\}
    \le \min\{4\newbeta\frac{z}{2^{j}},1\}\cdot4\newbeta\frac{d(u,v)}{2^j}$.
  \item $i=j-1$. For $\B_j$ to occur, the edge $e$ must be cut in round
    $j-1$ and must be untouched in round $j-2$. The probability of this
    is at most
    $\min\{\newbeta\frac{A_e}{2^{j-2}},1\}\cdot\newbeta\frac{d(u,v)}{2^{j-1}}
    \le \min\{4\newbeta\frac{z}{2^{j}},1\}\cdot2\newbeta\frac{d(u,v)}{2^j}$.
  \item $i=j$. For $\B_j$ to occur, $e$ must be cut in round $j$ and
    must be untouched in round $j-1$. The probability of this is at most
    $\min\{\newbeta\frac{A_e}{2^{j-1}},1\}\cdot\newbeta\frac{d(u,v)}{2^{j}}
    \le \min\{4\newbeta\frac{z}{2^{j}},1\}\cdot\newbeta\frac{d(u,v)}{2^j}$.
  \end{itemize}
  Since $\Pr[\B_i] = \Pr[\B_i \wedge (i\leq j-3)] +\Pr[\B_i \wedge
  (i=j-2)] + \Pr[\B_i \wedge
  (i=j-1)] +\Pr[\B_i \wedge (i=j)]$, the claim follows.
\end{proof}
\fi

\fullonly{
\subsection{\texorpdfstring{Connected \zex on General Graphs}{Connected 0-extension on General Graphs}}
\label{sec:ckr-conn}

Finally, we show that for general metrics, we can do
better than the $O(\log^2 k \log \log k)$ guarantee implied by
\lref[Theorem]{thm:connectedzextterminal}. In particular, we now prove
\lref[Theorem]{thm:connectedzextgen}, which gives a $O(\log k)$ guarantee.
We still use Algorithm~1 from the previous
section, but use a specific decomposition algorithm. The following
result follows from Fakcharoenpol et al.~\cite{FHRT03}, who built up on
the work of Calinescu, Karloff and Rabani~\cite{CKR04}:
\begin{theorem}[\cite{FHRT03}]
  \label{thm:ckrdecomp}
  Let $(G=(V,E), \ell_G)$ with a terminal set $K = \{x_1,\ldots,x_k\}
  \subseteq V$. There is a (randomized) polynomial-time algorithm that
  produces, for each $i = 0, 1, \ldots, \lceil \log \diam(G) \rceil$, a
  collection of $k+1$ clusters $\{C^i_0,C^i_1,\ldots,C^i_k\}$, such that
  \begin{OneLiners}
  \item[(a)] (Diameter) For any $j \neq 0$, $C^i_j$ contains the
    terminal $x_j$, and $d(x_j,v) \leq 2^i$ for any $v \in C^i_j$,
  \item[(b)] (Separation) For any $u,v\in X$, $\Pr[\mbox{$\exists j$
      such that $u \in C^i_j$ but $v\not\in C^i_j$}] \leq O(\beta^{uv}_i)\cdot
    d(u,v)/2^i,$ where the probability is taken over the internal coin
    tosses of the algorithm, and
  \item[(c)] (Amortization) For any $u,v \in X$, $\sum_{i} \beta^{uv}_i \leq \beta = O(\log k)$.
  \item[(d)] (Coverage) $\cup_{j \neq 0} C^i_j$ contains $\cup_{j=1}^k
    B_d(x_j, 2^{i-1}).$
  \end{OneLiners}
\end{theorem}

We remark that we do not need each cluster to induce a connected component.
Observe that the (Diameter) and (Coverage) properties imply:
{\em
\begin{OneLiners}
\item[(e)] (Laminarity) For any $i$, $\cup_{j \neq 0} C^{i+1}_j \supseteq
  \cup_{j \neq 0} C^i_j$ with probability $1$. Hence also $C^i_0
  \supseteq C^{i+1}_0$ with probability $1$.
\end{OneLiners}
}

We run Algorithm~1 with this decomposition; the only worry is that since
the clusters are not connected, it may be the case that in
step~\ref{line:wnonsteiner}, we may not find a node $w_C$ as desired.
In this case, we {\em expel} $C$ from $C_s$, and do not map the vertices
in $C$ in this iteration. This ensures the connectivity property of
$f_{i}^{-1}(x)$'s. Moreover, the Laminarity property inductively ensures
that we never map any vertex from $C^i_0$ by the end of round $i$. Since
the diameter property bounds the diameter of every other cluster,
\lref[Lemma]{lem:smallclusters} continues to
hold. 

Now, by its very definition, any expulsion operation only removes
components that are disconnected from the rest of $C_s$, and hence does
not increase the separation probability for any edge. Moreover, it is still the case the if $u$ is mapped before round $j$ and an edge $(u,v)$ is not cut in round $j$, then the node $v$ gets mapped in round $j$ as well. Indeed by laminarity, $u$ is in one of the clusters containing a terminal, and if $(u,v)$ is not cut, then so is $v$.  Since $u$ is mapped, the component containing $v$ cannot be expelled. Thus \lref[Claim]{cla:probxj} continues to hold and bounds the probability of $\B_j$, implying that 
\begin{equation*}
\begin{split}
\ts \EX[d(f(u),f(v))] &= \ts \sum_{j}\Pr[\B_j]\cdot \EX[d(f(u),f(v))\mid \B_j]\\
& \ts \le O(d_G(u,v)) + \sum_{j\ge j'}\Pr[\B_j]\cdot O(2^j)\\
&\ts \le O(d_G(u,v)) +
\sum_{j\ge
  j'}\min\{{\textstyle4\beta\frac{z}{2^j}},1\}\cdot{\textstyle5\beta^{uv}_i\frac{d_G(u,v)}{2^j}}
\cdot O(2^j) \quad \le O(\beta\,d_G(u,v)).\enspace
\end{split}
\end{equation*}
Since $\beta = O(\log k)$, this gives us connected {\zex}s where the stretch
is $O(\log k)$, and hence finishes the proof of
\lref[Theorem]{thm:connectedzextgen}.
}

\section{Lower Bounds}
\label{sec:lower-bounds}

In this section, we show two kinds of lower bounds. The first shows that
any flow-sparsifier that is a convex combination of {\zex}s must suffer a
loss of $\Omega(\sqrt{\log k})$---for such extension, this improves on the
$\Omega(\log\log n)$ lower bound for (arbitrary) flow-sparsifiers~\cite{LM10}. 
The second shows that any flow-sparsifier that
only uses edge capacities which are bounded from below by a constant, must
suffer a loss of $\Omega(\sqrt{\log k}/\log\log k)$.

\subsection{Lower Bounds for 0-Extension-Based Sparsifiers}
\label{sec:lbd-zex}

The following result can be viewed as following from the duality between {\zex}s and \zex-based flow-sparsifiers (Theorem~\ref{thm:af-thm}); by that theorem, not only do good \zex algorithms give good \zex-based flow-sparsifiers, the converse would also be true---and hence one can use a lower bound of Calinescu et al.~\cite{CKR04} to infer lower bounds on \zex-based flow-sparsifiers. The following theorem gives the explicit construction obtained thus.

\begin{theorem}
\label{thm:zex-lbd}
For infinitely many values of $k$, there is a graph $G'=(V(G'),E(G'))$ and a set $K\subseteq V$ of size $k$ for which any flow-sparsifier that is a convex combination of 0-extension graphs has quality at least $\Omega(\sqrt{\log k})$.
\end{theorem}

\begin{proof}
We use the lower bound of $\Omega(\sqrt{\log k})$ on the 0-extension integrality ratio by Calinescu et al.~\cite{CKR04}. For completeness we describe their construction: Let $G$ be an expander with $n$ vertices, maximum degree $\Delta$ and expansion at least $\alpha$, where $\Delta$ and $\alpha$ are fixed parameters. Define $l=\left\lceil \sqrt{\log n}\right\rceil $ and $k=\left\lceil \frac{n}{l}\right\rceil $. Choose any $k$ distinct vertices $h_{1},\dots.h_{k}\in V\left(G\right)$ and add $k$ new paths of length $l$ starting at these vertices and ending at new vertices labeled $1,\dots,k$. Denote the resulting graph by $G'$ (note that $\left|V\left(G'\right)\right|=O\left(n\right)$ and $\left|E\left(G'\right)\right|=O\left(n\right)$), and let the terminals $K$ be the new vertices $\left\{ 1,\dots,k\right\} $.  Set the costs and lengths of the edges to 1. The distance $d_{G'}\left(u,v\right)$ is set to be the shortest path distance in $G'$ between $u,v$. For the described instance $G',K$ of the 0-extension problem, Calinescu et al.$\ $show that
\begin{align*}
\sum_{e=\left(u,v\right)\in E\left(G'\right)} c(e) d_{G'}(u,v) = \left|E\left(G'\right)\right| = O(n),
\end{align*}
while there exists a universal $\gamma>0$ such that for any 0-extension function $f:V\left(G'\right)\rightarrow K$,
\[
\sum_{e=(u,v)\in E(G')} c(e) d_{G'} (f(u),f(v)) \ge \gamma n\sqrt{\log n} = \Omega\left(n\sqrt{\log k}\right).
\]

We now use the instance $G',K$ as follows. By \cite[proof of Theorem 1]{LM10} it is known that for any convex combination of 0-extensions $H=\sum\lambda_{i}H_{i}$, the quality of $H$ is 
\begin{eqnarray*}
\sup_{d_{G'} \ \text{s.t.} \ \sum_{e} c(e)d_{G'}(e)=1} &\{\sum_{s,t\in K} c_H(s,t)d_{G'}(s,t)\} & = \\
\sup_{d_{G'} \ \text{s.t.} \ \sum_{e} c(e)d_{G'}(e)=1} & 
\{ \sum_{f_i} \lambda_i \sum_{(u,v)\in E(G')} c(e) d_{G'}(f_i(u),f_i(v))\} &.
\end{eqnarray*}
(The proof of this uses strong duality for the maximum concurrent flow problem.) We now show that there exists a semimetric $d_{G'}$ such that $\sum_{e} c(e)d_{G'}(e)=1$, and for every 0-extension function $f:V\left(G'\right)\rightarrow K$,
\begin{eqnarray}
\sum_{(u,v)\in E(G')} {c(e)d_{G'}(f(u),f(v))} = \Omega\left(\sqrt{\log k}\right).\label{eq:LB}
\end{eqnarray}

We set $d_{G'}(e)$ to be $1/|E(G')|$ for every $e\in E(G')$. Thus, $\sum_{e\in E(G')} c(e)d_{G'}(e) = 1$. We set $d_{G'}(u,v)$ to be the shortest path distance between $u,v$ in $G'$ with respect to edge lengths $d_{G'}(e)$. From above it follows that every 0-extension function $f$,
\begin{eqnarray*}
\sum_{(u,v)\in E(G')} {c(e)d_{G'}(f(u),f(v))} \ge \frac{\gamma n\sqrt{\log n}} {\left|E\left(G'\right)\right|} = \Omega\left(\sqrt{\log n}\right) = \Omega\left(\sqrt{\log k}\right).
\end{eqnarray*}

This proves Equation \ref{eq:LB}, completing the proof.
\end{proof}

\subsection{Lower Bounds for Sparsifiers having no Small Edges}
\label{sec:lbd-small}

\begin{theorem}
\label{thm:sparsifier-lbd}
 For infinitely many values of $k$, there is a graph
  $G=\left(V,E\right)$ and a terminal set $K\subset V$ of size $k$ for which any
  flow-sparsifier with edge capacities at least $\varepsilon>0$ has
  quality at least $\Omega(\varepsilon\sqrt{\log k}/\log\log k)$.
\end{theorem}
\begin{proof}
Let $n$ be a sufficiently large prime. Let $G=(V,E)$ be a graph whose nodes
correspond to the elements of $\mathbb{Z}_n$ and that contains an edge $\{u,v\}$
if $v=u+1$, $v=u-1$, or $v=u^{-1}$ (all operations are w.r.t. $\mathbb{Z}_n$ and
we define $0^{-1}$ as 0.)
In other words the graph consists of a Hamiltonian cycle plus some additional edges.
This graph $G$ is a 3-regular expander (see, e.g.,~\cite{HLW06}).

Choose the set of terminals $K$ as $\{i\cdot \lceil\sqrt{\log n}\rceil \mid 0\le i\le k-1\}$,
with $k= n/\lceil\sqrt{\log n}\rceil$. To simplify notation, we will omit floor- and
ceiling-operations in the following.
For $i\in[0,k-1]$, let $B_i$ be the set of the $\sqrt{\log n}$ nodes on the Hamiltonian
cycle between terminal $i$ and $i+1$, including $i$ but excluding $i+1$.

Let $H=(K,E_H)$ be a flow-sparsifier for $G$ with edge capacities at least
$\varepsilon>0$. Let $d$ be the maximum weight degree of $H$, where the
weighted degree of a node is the sum over all capacities of incident edges.
\begin{claim}\label{clm:max_weight_degree}
The maximum weighted degree $d$ of $H$ is at least
\[c' \cdot\varepsilon\cdot \frac{\sqrt{\log n}}{\log\log n} \]
for some constant $c'$.
\end{claim}
\begin{proof}
Consider a demand of $1/k$ between all pairs of terminals.

Since the minimum edge capacity is at least $\varepsilon$, the unweighted
degree of $H$ is at most $d/\varepsilon$. Due to this bounded degree, for
sufficiently large $k$, there are at least $k^2/4$ terminal pairs that have
distance at least $\log k/(2\log(d/\varepsilon))$ from each other (see e.g.
\cite[Lemma 4.2]{CKR04}).

Each of these pairs induces a load of $1/k$ on at least $\log
k/(2\log(d/\varepsilon))$ edges. Therefore, the total load in the network is at
least $k\log k/(8\log(d/\varepsilon))$. Since $H$ has at most $k\cdot
d/(2\varepsilon)$ edges, the congestion in $H$ is at least $\varepsilon\log
k/(4 d\log(d/\varepsilon))$.

The same demand can be routed with congestion at most $(c+1)\sqrt{\log n}$ in $G$,
for some constant $c$ depending on the edge expansion of $G$.
Say each terminal $i$ sends a total flow of 1. We can distribute this flow
evenly between the nodes in $B_i$ using only edges inside of $B_i$ and with
congestion of at most $1$. This can easily be done, since we can send
this flow along the Hamiltonian cycle to reach every node in $B_i$.
Now, we route a uniform multicommodity flow on the whole expander, where the
flow leaving each node is $1/\sqrt{\log n}$, i.e., the demand between every
pair of nodes is $1/(n\sqrt{\log n})$. This requires congestion at most
$c\log n \cdot (1/\sqrt{\log n}) = c\sqrt{\log n}$~\cite{LR99}. Finally, the flow in each $B_i$ is routed inside
$B_i$ to the respective terminal. Again, this can easily be done with
congestion $1$. In total, we sent a flow of $1/k$ between all pairs of
terminals and the congestion is bounded by $c\sqrt{\log n}+2\le (c+1)\sqrt{\log
  n}$.

Hence, we identified a demand, that requires congestion at least
$\varepsilon\log k/(4 d\log(d/\varepsilon))$ in $H$ but can be routed with
congestion at most $(c+1)\sqrt{\log n}$ in $G$. Since $H$ is a flow-sparsifier,
its congestion has to be bounded by the congestion in $G$ and thus,
$\varepsilon\log k/(4 d\log(d/\varepsilon)) \leq (c+1)\sqrt{\log n}$. It
follows that
\[
\frac{d}{\varepsilon}\log\Big(\frac{d}{\varepsilon}\Big)\ge \frac{\log
k}{4(c+1)\sqrt{\log n}} \enspace.
\]
Using the fact that $k=n/\sqrt{\log n}$, the claim follows.
\end{proof}

Now pick a node in $H$ that as weighted degree at least
$c'\cdot\varepsilon\cdot\sqrt{\log n}/\log\log n$ (such a nodes exists due to
Claim~\ref{clm:max_weight_degree}). Consider the situation in which the demand
between this node and every other node corresponds to the capacity of the edge
connecting them in $H$, and all other demands are 0. Clearly, in $H$ this can
be routed with congestion 1. The terminal in $G$ corresponding to node $u$,
however, has only degree $3$. Therefore, routing this demand in $G$ results in
congestion at least $c'\cdot\varepsilon\cdot\sqrt{\log n}/(3\log\log n)\ge
c'\cdot\varepsilon\cdot\sqrt{\log k}/(3\log\log k)$, since that is the load on
at least one of the outgoing edges of $u$.
\end{proof}

\section{Applications}
\label{sec:applications}

Most of these applications were considered by Moitra~\cite{Moitra09},
and Leighton and Moitra~\cite{LM10}; we show how our results above give
improved approximations to the problems.

\subsection{Steiner Oblivious Routing}

\lref[Theorem]{thm:cong-restate} is an exact analogue of R\"acke's
theorem on general flows~\cite{R08} for the special case of $K$-flows,
and hence immediately gives an $O(\log k)$-oblivious routing scheme for
$K$-flows.

\subsection{Steiner Minimum Linear Arrangement}

Given $G = (V,E)$ and $K \sse V$ with $|K| = k$, the goal in the Steiner
Minimum Linear Arrangement (SMLA) problem is to find a mapping $F: V \to
[k]$ such that $F|_K: K \to [k]$ is a bijection. The goal is to minimize
$\sum_{(u,v) \in E} c_{uv} |F(u) - F(v)|$. Note that for the non-Steiner
MLA case where $K = V$, Rao and Richa~\cite{RaoR98} gave an $O(\log
n)$-approximation for general graphs and an $O(\log \log
n)$-approximation for graphs that admit $O(1)$-padded decompositions
(which includes the family of all trees).

For our algorithm, we take a random tree/retraction pair $(T, f)$ from
the distribution of Theorem~\ref{thm:logk-GNR}; this ensures that the cost of the optimal map
$F^*$ (viewed as a solution to the MLA problem on $T$) increases by an
expected $O(\log k)$-factor. Now solving the MLA problem on the tree to
within an $O(\log \log k)$ factor to get a map $\widehat{F}_T : K \to
[k]$, and defining $\widehat{F}(x) = \widehat{F}_T(f(x))$ gives us an
expected $O(\log k \log\log k)$-approximation. We show in \lref[Section]{sec:direct-algos} that this can be improved slightly to $O(\log k)$ using a more direct approach.

\subsection{Steiner Graph Bisection}

In this problem, we are given a value $k'$ and want to find a
bipartition $(A, V\setminus A)$ of the graph such that $|A \cap K| =
k'$, and that minimizes the cost of edges cut by the bipartition. We use
Theorem~\ref{thm:cong-restate} to embed the graph into a random tree
losing an $O(\log k)$ factor. On this tree we use 
the
approach of R\"acke~\cite{R08} to find the best $(k', k-k')$ bipartition
on that. This gives us an $O(\log k)$
algorithm for this partitioning problem.

\subsection{\texorpdfstring{Steiner $\ell$-Multicut}{Steiner l-Multicut}}

In this problem, we are given terminal pairs $\{s_i, t_i\}_{i \in [k]}$,
and a value $k' \leq k$, and we want to find a minimum cost set of edges
whose deletion separates at least $k'$ terminal pairs. Again, we can
use Theorem~\ref{thm:cong-restate} to embed the graph into a random tree losing an $O(\log k)$ factor, and use
the theorem of Golovin et al.~\cite{GNS06} to get a $4/3 +
\epsilon$-approximation on this tree; this gives us the randomized
$O(\log k)$-approximation.

\subsection{Steiner Min-Cut Linear Arrangement}

The Steiner Min cut Linear Arrangement (SMCLA) problem is defined as follows: Given $G = (V,E)$ and $K \sse V$ with $|K| = k$, we want to find a mapping $F: V \to [k]$ such that $F|_K: K \to [k]$ is a bijection. The goal is to minimize $\max_{i} \sum_{x \in F^{-1}([i]), y \not \in F^{-1}([i])} c_{xy} $. For the non-Steiner version of the problem, Leighton and Rao~\cite{LR99} show that given an $\alpha$-approximation to the balanced partitioning (or to the bisection) problem, one can get an $O(\alpha\log n)$-approximation to the MCLA problem. Using~\cite{AroraRV09}, this gives an $O(\log^{1.5} n)$-approximation to the MCLA problem.

We note that the reduction works immediately for the Steiner version of the problem: given an $\alpha$-approximation to Steiner-bisection, one gets an $O(\alpha \log k)$-approximation to SMCLA. Thus we get an $O(\log^2 k)$-approximation to the SMCLA problem. We show in \lref[Section]{sec:direct-algos} that this can be improved to $O(\log^{1.5} k)$ using a more direct approach.

\section{Better Algorithms Using a Direct Approach}
\label{sec:direct-algos}

The vertex-sparsifiers give a modular approach to solving steiner
version of various problems. Not surprisingly, for some of these
problems, a direct attack will lead to better algorithms. In this
section, we show that applying known techniques for Minimum Linear
Arrangement (MLA) problem lead to a better approximation ratio for
Steiner MLA, and for Steiner Minimum Cut Linear Arrangement.

\subsection{Steiner Minimum Linear Arrangement}

Recall that the Steiner MLA problems is defined as follows. Given $G =
(V,E)$ and $K \sse V$ with $|K| = k$, the goal is to find a mapping $F:
V \to [k]$ such that $F|_K: K \to [k]$ is a bijection. The goal is to
minimize $\sum_{(u,v) \in E} c_{uv} |F(u) - F(v)|$. Specifically, we
show the following result.
\begin{theorem}
  \label{thm:smlalp}
  There is a polynomial time $O(\log k)$-approximation algorithm for the
  SMLA problem based on the natural linear programming relaxation.
\end{theorem}

\begin{proof}
The linear program for the SMLA problem is based on the spreading metric
linear programming relaxation for MLA introduced in~\cite{EvenNRS00}.
\[
\begin{array}{rrcll}
\min & \sum_{(u,v) \in E} c_{uv} d_{uv}\\
\mbox{subject to:} &\\
\mbox{(Triangle Inequality)}& d_{uw} - d_{uv} - d_{vw}& \le & 0 & \forall u,v,w \in V\\
\mbox{(Spreading)} & \sum_{v\in S}d_{uv} & \ge & \frac{|S|^2}{5} & \forall S \subseteq K, |S| \geq 2, u \in S\\
 & d_{uv} & \ge & 0 & \forall u,v \in V\\
\end{array}\]

It follows from~\cite{EvenNRS00} that the above is a valid linear
programming relaxation to the SMLA problem, and that one can efficiently
separate for the spreading constraints so that the LP can be solved in
polynomial time using the Ellipsoid algorithm. Further, it is easy to
check that the spreading constraints imply that for any $u \in K$,
$|\mathbf{B}_d(u,r) \cap K| \leq 5r$. (Here, $\mathbf{B}_d(v, r) = \{ w \mid
d(v,w) \leq r\}$ is the ``ball'' around $v$ of radius $r$ in the metric
$d$.)

Let $d$ be a solution to the above linear program. Since $d$ is a metric
on $V$, it follows from \lref[Theorem]{thm:logk-GNR} that we construct a
(random) edge-weighted $2$-HST $T = (I \cup K, E_T)$ with internal nodes
$I$ and leaves $K$, and a retraction $f: V \to K$ such that
\begin{OneLiners}
  \item[(a)] $d_T(f(x),f(y)) \geq d(x,y)$ for all $x, y \in K$ (with
    probability $1$),
  \item[(b)] $\EX_T[ d_T(f(u),f(v)) ] \leq O(\log k) \; d(u,v)$ for all
    $u, v \in V$.
\end{OneLiners}

We argue that given this HST, we can construct a mapping $F_T: V \to
[k]$ such that $F_T|_K: K \to [k]$ is a bijection. This mapping will
have the property that $|F_T(u)-F_T(v)| \leq 5 d_T(f(u),f(v))$. The
approximation ratio of $O(\log k)$ then follows from property~(b) above.

The mapping $F_T$ is defined by taking the natural left-to-right
ordering on $K$ defined by $T$, and assigning every other vertex $v \in
V$ to the position $f(v)$. Formally, let $\pi$ be a pre-order traversal
of $T$.  For every terminal $x \in K$, set $F_T(x)$ to the number of
terminals in $\pi$ that occur before $x$, i.e. $F_T(x) = |K \cap \{\pi_i
: i \leq \pi^{-1}(x)\}|$. For every other vertex $u \in V$, set $F_T(u)=
F_T(f(u))$. It is easy to check that $F_T|_K$ is a bijection.

We next upper bound $|F_T(u)-F_T(v)|$ for $u,v \in V$. Consider the
terminals $t_u = f(u), t_v = f(v)$; if $t_u=t_v$, then $F_T(x)=F_T(y)$
and there is nothing to prove. Else let $T_{uv}$ be the smallest subtree
of $T$ containing $t_u$ and $t_v$.  By the properties of the HST, we
have $d_T(t_u,t_v) \geq d_T(t_u,z)$ for all $z \in T_{xy}$. Moreover,
$d_T(u,v) = d_T(t_u,t_v)$. Now,
\begin{align}
  |F_T(u)-F_T(v)| &= |F_T(t_u)-F_T(t_v)| \notag \\
  &\leq |K \cap T_{uv}| \notag \\
  & \leq  |K \cap \mathbf{B}_{d_T}(t_u, d_T(t_u,t_v))| \tag{\text{Since
      $d_T(t_u,t_v) \geq d_T(t_u,z)$ for all $z \in T_{uv}$}} \\
  & \leq  |K \cap \mathbf{B}_{d}(t_u, d_T(t_u,t_v))| \tag{\text{By
      property~(a)}} \\
  &\leq  5  d_T(t_u,t_v) \tag{\text{By the spreading property}}\\
  &= 5d_T(u,v). \notag
\end{align}
This proves
\lref[Theorem]{thm:smlalp}
  \end{proof}

\subsection{Steiner Min Cut Linear Arrangement}

Recall that the Steiner Min cut Linear Arrangement (SMCLA) problem is defined as follows. Given $G = (V,E)$ and $K \sse V$ with $|K| = k$, the goal is to find a mapping $F: V \to [k]$ such that $F|_K: K \to [k]$ is a bijection. The goal is to minimize $\max_{i} \sum_{x \in F^{-1}([i]), y \not \in F^{-1}([i])} c_{xy} $. Specifically, we show the following result.
\begin{theorem}
\label{thm:smclalp}
There is a polynomial time $O(\log^{1.5} k)$-approximation algorithm for the SMCLA problem.
\end{theorem}

The algorithm and the proof are the natural generalization of the $O(\log^{1.5} n)$ approximation to the min cut linear arrangement problem. We sketch the argument here.

This algorithm is based on an SDP formulation and the sparsest cut algorithm of ~\cite{AroraRV09}, who show the following theorem. 

\begin{theorem}
\label{thm:arv}
There exist a constant $\eps>0$ such that the following holds. For any $k$-point $\ell_2^2$ metric $(S,d)$ satisfying $\sum_{x,y \in S} d_{xy} \geq \frac{|S|^2}{8}$, there are sets $A,B \subseteq S$ such that $|A|,|B| \geq \eps k$ and $d(A,B) \geq \frac{\eps}{\sqrt{\log k}}$. Moreover given vectors $\{v_x : x\in S\}$ representing $d$, such sets $A,B$ can be found in polynomial time.
\end{theorem}

Consider first the following linear program:
\[
\begin{array}{rrcll}
\min & \sum_{(x,y) \in E} c_{xy} d_{xy}\\
\mbox{subject to:} &\\
\mbox{(Triangle Inequality)}& d_{xz} - d_{xy} - d_{yz}& \le & 0 & \forall x,y,z \in V\\
\mbox{(Balance)} & \sum_{x,y\in K }d_{xy} & \ge & \frac{|K|^2}{8} & \\
 & d_{xy} & \ge & 0 & \forall x,y \in V\\
\end{array}\]

Let $F: V \to [k]$ be the optimum MCSLA with value $OPT$. Then the cut separating $F^{-1}([\lfloor \frac{k}{2}\rfloor])$ from its complement has value at most OPT, and gives a feasible integral solution to above linear program. Thus the value of the relaxation above is at most $OPT$.

Suppose in the above linear program, we additionally require that the distance metric $d$ be an $\ell_2^2$ metric, i.e. there exists vectors $v_x \in \R^n$ such that $d(x,y) = \|v_x-v_y\|_2^2$. This program can be naturally written as an SDP, and can be solved in polynomial time to return vectors $\{v_x\}$. Moreover, the optimum to this relaxation has value at most $OPT$ as well. Theorem~\ref{thm:arv} then implies that we can find sets $A, B \subseteq K$ such that $|A|,|B| \geq \eps k$ and where $d(A,B) \geq \Delta = \frac{\eps}{\sqrt{\log k}}$. Consider the sets $A_r = \{x \in V: d(A,x) \leq r\}$. For $0< r < \Delta$, it is immediate that $A \subseteq A_r \subseteq V\setminus B$.

Picking $r$ at random from $(0,\Delta)$, we observe that for any $x,y\in V$
$$\Pr[x \in A_r, y \not \in A_r] \leq (d(y,A)-d(x,A))/\Delta,$$
so that by triangle inequality, the expected cost of the cut $(A_r,V\setminus A_r)$ is at most $\frac{1}{\Delta} \sum_{(x,y) \in E} c_{xy} d_{xy} \leq OPT/\Delta$. Thus we can find an $r \in (0,\Delta)$ such that
\begin{OneLiners}
\item[(a)] $|K \cap A_r|, |K \cap (V \setminus A_r)| \leq (1-\eps)k.$
\item[(b)] $\sum_{x \in A_r, y \not\in A_r} c_{xy} \leq O(OPT\sqrt{\log k}).$
\end{OneLiners}

We can recursively compute steiner linear arrangements for $A_r$ and $V\setminus A_r$, and by condition $(a)$, the depth of the recursion is at most $O(\log k)$. For any $i$, we can thus bound the total cost of edges from $F^{-1}([i])$ to $V \setminus F^{-1}([i])$. Indeed each level of the recursion contributes at most $O(OPT \sqrt{\log k})$ to this cost. Since there are at most $O(\log k)$ levels, we get an $O(\log^{1.5} k)$ approximation.

\ifprocs
\else

\fi

\procsonly{\vspace{-0.05in}}
{\small
\fullonly{\bibliography{bib}}
\procsonly{\bibliography{bibabbrev}}
}

\end{document}